\documentclass[a4paper,11pt]{article}
 \pdfoutput=1
 \usepackage{jheppub} 
 \usepackage{graphicx}
\usepackage{float}
 \usepackage{amssymb}
\usepackage{dcolumn}% Align table columns on decimal point
\usepackage{bm}% bold math 
\usepackage{color}
\usepackage{multirow}

\allowdisplaybreaks

 \newcommand{\lsim}{{\;\raise0.3ex\hbox{$<$\kern-0.75em\raise-1.1ex\hbox{$\sim$}}\;}}
\newcommand{\gsim}{{\;\raise0.3ex\hbox{$>$\kern-0.75em\raise-1.1ex\hbox{$\sim$}}\;}}
\newcommand{\beq}{\begin{equation}}
\newcommand{\eeq}{\end{equation}}
\newcommand{\bea}{\begin{eqnarray}}
\newcommand{\eea}{\end{eqnarray}}
\mathchardef\minus="002D

\newcommand{\mev}{\textrm{ MeV}} 
\newcommand{\gev}{\textrm{ GeV}}

\newcommand{\eps}{\varepsilon}
\newcommand{\br}{{\rm BR}}
\newcommand{\nn}{\nonumber}

\def\beq{\begin{equation}}
\def\eeq{\end{equation}}
\def\bea{\begin{eqnarray}}
\def\eea{\end{eqnarray}}
\def\bit{\begin{itemize}}
\def\eit{\end{itemize}}

\def\baa{\begin{array}}
\def\eaa{\end{array}}

\def\met{E\hspace{-0.25cm}/_{T}}

\title{\boldmath Invisible dark gauge boson search in top decays using a kinematic method}

\author[a]{Doojin Kim}   
\author[b]{Hye-Sung Lee} 
\author[c,d,e]{Myeonghun Park} 

\affiliation[a]{Physics Department, University of Florida, Gainesville, FL 32611, USA}
\affiliation[b]{Theory Division, CERN, CH-1211 Geneva 23, Switzerland}
\affiliation[c]{Asia Pacific Center for Theoretical Physics, 77 Cheongam-Ro, Nam-Gu, Pohang 790-784, Korea}
\affiliation[d]{Department of Physics, Postech, Pohang 790-784, Korea}
\affiliation[e]{Kavli IPMU (WPI), The University of Tokyo, Kashiwa, Chiba 277-8583, Japan}
\preprint{
\begin{minipage}[r]{2.in} 
\begin{flushright}
CERN-PH-TH-2014-209 \\
APCTP Pre2014-013\\
IPMU14-0330
\end{flushright}
\end{minipage}
}

\emailAdd{immworry@ufl.edu}
\emailAdd{hyesung.lee@cern.ch}
\emailAdd{parc@apctp.org}

\abstract{We discuss the discovery potential of a dark force carrier ($Z'$) of very light mass, $m_{Z'} \lsim {\cal O}(1-10) \gev$, at hadron colliders via rare top quark decays, especially when it decays invisibly in typical search schemes.
We emphasize that the top sector is promising for the discovery of new particles because top quark pairs are copiously produced at the Large Hadron Collider.
The signal process is initiated by a rare top decay into a bottom quark and a charged Higgs boson ($H^\pm$) decaying subsequently into a $W$ and one or multiple $Z'$s.
The light $Z'$ can be invisible in collider searches in various scenarios, and it would be hard to distinguish the relevant collider signature from the regular $t\bar{t}$ process in the Standard Model.
We suggest a search strategy using the recently proposed on-shell constrained $M_2$ variables.
Our signal process is featured by an {\it asymmetric} event topology, while the $t\bar{t}$ is {\it symmetric}. The essence behind the strategy is to evoke some contradiction in the relevant observables by applying the kinematic variables designed under the assumption of the $t\bar{t}$ event topology. 
To see the viability of the proposed technique, we perform Monte Carlo simulations including realistic effects such as cuts, backgrounds, detector resolution, and so on at the LHC of $\sqrt{s}=14$ TeV. }

\begin{document}
\maketitle
\flushbottom

%%%%%%%%%%%%%%%%%%%%%%%%%%%%%%%%%%%%%%%%%
%%%%%%%%%%%%%%%%%%%%%%%%%%%%%%%%%%%%%%%%%
\section{\label{sec:Intro} Introduction}
%%%%%%%%%%%%%%%%%%%%%%%%%%%%%%%%%%%%%%%%%
%%%%%%%%%%%%%%%%%%%%%%%%%%%%%%%%%%%%%%%%%
The Standard Model (SM) has been remarkably successful in explaining a wide range of phenomena in nature with high accuracy. Moreover, the discovery of a new scalar state at the Large Hadron Collider (LHC)~\cite{Aad:2012tfa,Chatrchyan:2012ufa}, which is consistent with the SM Higgs boson, reaffirms the role of the SM as a proper description of fundamental particles. Nevertheless, there still exist phenomena that cannot be explained by the SM. One definite example is the dark matter (DM), which is originally rooted in astrophysical observations~\cite{Bertone:2004pz}.

Dark force has been paid attention largely because of its potential link to the dark sector where the DM belongs to. It is considered to be a hypothetical interaction among the particles in the dark sector. In particular, if those particles do not couple to any of the known forces but communicate with the SM sector via the relevant dark force carrier coupled to some SM force carrier, then the existence of  
dark force becomes of paramount importance for exploiting the dark sector. In regards of astrophysical anomalies such as positron excess reported by various cosmic ray experiments including the PAMELA~\cite{Adriani:2008zr}, the Fermi Gamma-ray Space Telescope~\cite{FermiLAT:2011ab}, and the AMS-02~\cite{Aguilar:2013qda}, dark force furnishes with a theoretical basis to explain those phenomena. Furthermore, for the light DM candidate, which has been recently reported by the CDMS experiment~\cite{Agnese:2013rvf}, it also provides a consistent picture with their observation. There exist many other phenomena that motivate a new force carrier including the muon anomalous magnetic moment~\cite{Fayet:1980rr,Gninenko:2001hx,Fayet:2007ua,Pospelov:2008zw}. We also refer to the latest Snowmass report~\cite{Essig:2013lka} and the references therein for extensive discussions on the theoretical and observational motivations of dark force. 

Given such appealing motivations, many new physics models introducing dark force mediators such as the dark photon \cite{ArkaniHamed:2008qn} and the dark $Z$ \cite{Davoudiasl:2012ag} have been proposed, and at the same time active searches for them are underway. We shall call such dark gauge bosons $Z'$ throughout this paper, independent of the model. The typical mass of dark force carriers is roughly of GeV scale, and therefore, both low and high energy experiments can search for them.
In the low energy experiments, the $Z'$ production typically depends on the bremsstrahlung and meson decays for the $Z'$ production \cite{Bjorken:2009mm,Essig:2013lka}.
In the high energy experiments, decays from heavy particles such as a Higgs boson \cite{Davoudiasl:2012ig,Lee:2013fda,Davoudiasl:2013aya,Davoudiasl:2014mqa} can be exploited for the light $Z'$ production.
There are also studies in the supersymmetry context \cite{ArkaniHamed:2008qp}.
In this paper, we emphasize the usefulness of the top sector. Since the top quark, which is the heaviest among the known elementary particles, is expected to be copiously produced at the LHC, the top sector can provide great channels for the light $Z'$ search, in conjunction with the relatively poor measurement of the top quark decays.
We remark that typical errors in the top quark decays are of ${\cal O}(10 \%)$ \cite{Agashe:2014kda}.

%%%%%%%%% FIGURE %%%%%%%%
\begin{figure}[t]
\centering
\includegraphics[width=6.6cm]{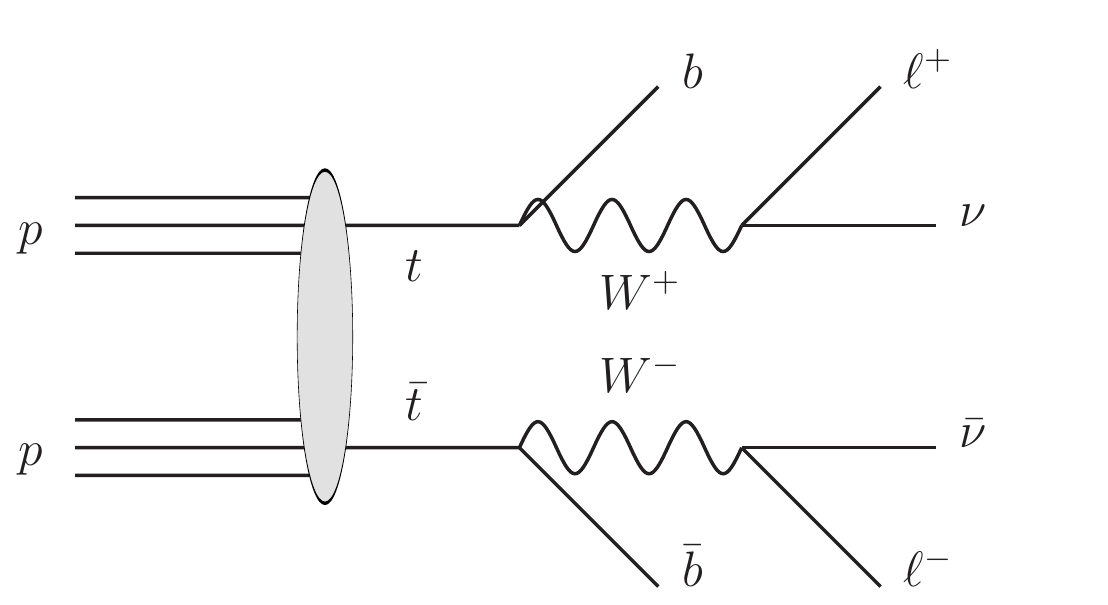} \qquad
\includegraphics[width=7.4cm]{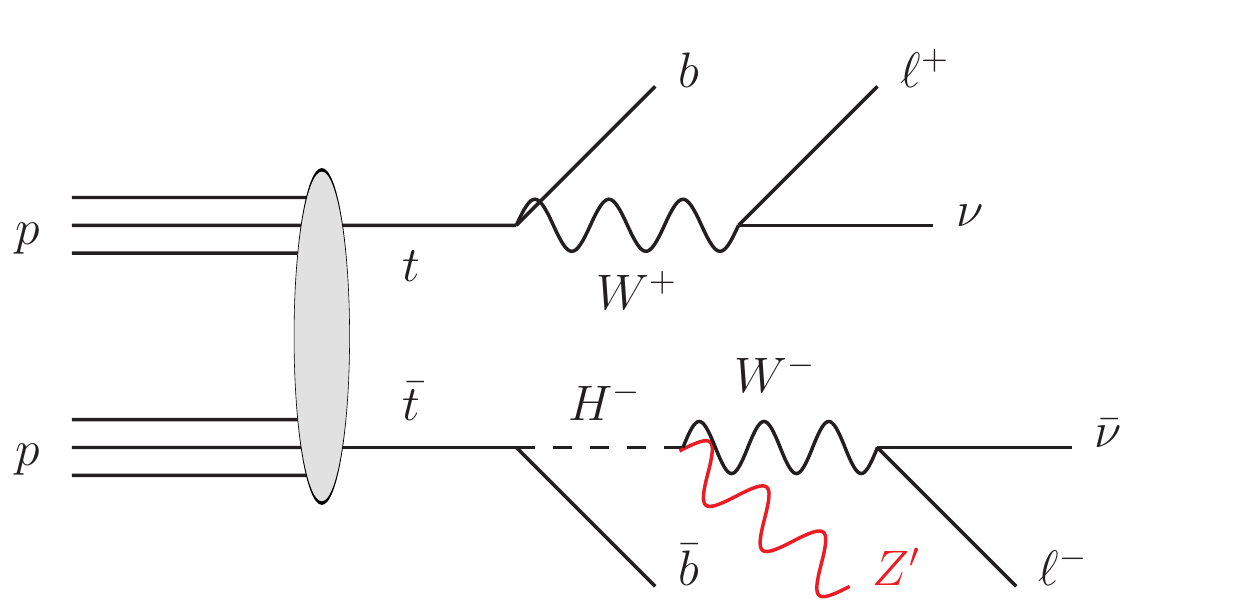}
\caption{\label{fig:decayTopo} The background (left panel) and signal (right panel) decay topologies. They are the same except for the new light gauge boson $Z'$, which is invisible in various scenarios.}
\end{figure}
%%%%%%%%%%%%%%%%%%%%%%

Recently, it has been pointed out that the light $Z'$ can be produced as a dominant decay product of a charged Higgs boson in the relevant models~\cite{Lee:2013fda,Davoudiasl:2014mqa}.
The decay of top quark can be a good channel to probe such a dark force \cite{Davoudiasl:2014mqa,Kong:2014jwa}, and a study of the lepton-jet ($Z' \to \ell^+ \ell^-$ with highly collimated leptons) signature of the $Z'$ via $t \to b H^{\pm} \to b W Z'$ in the LHC $t \bar t$ channel was performed in Ref.~\cite{Kong:2014jwa}.
In this paper, we perform a complementary study for the models with the invisible $Z'$ which can actually cover, at least, three different cases enumerated below:
\begin{itemize}
\item[(1)] $Z'$ decaying dominantly into very light dark sector particles $\chi$, e.g., $Z' \to \chi \bar\chi$, 
\item[(2)] $Z'$ decaying into dileptons or dijets, which can escape in most analysis schemes due to their high collimation,
\item[(3)] $Z'$ being extremely light (e.g., $m_{Z'} < 1 \mev$) so that its decay into visible particles is kinematically closed.
\end{itemize}

Case (1) assumes the presence of dark matter, which is even lighter than the light $Z'$.
The interest in this type has been growing partly because the visibly-decaying dark photon (a popular dark force model) has been excluded as a solution to the $g_\mu - 2$ anomaly by the recent experiments \cite{Adare:2014mgk}, while the invisibly-decaying dark photon still stands as a possible solution (see the discussions in Ref.~\cite{Lee:2014tba}). 
Case (2) becomes relevant along with the fact that because the $Z'$ easily gets boosted due to its light mass, the dileptons or dijets are highly collimated. Dijets are hard to separate in the hadron collider backgrounds, while the dileptons require the dedicated lepton-jet search (for instance, see Ref.~\cite{Kong:2014jwa}) as the typical isolated lepton cuts fail for such collimated leptons. Even the lepton-jet search may not be successful for very light $Z'$ due to the increasing background from $\gamma^* \to \ell^+ \ell^-$ \cite{Kong:2014jwa}. More generally, Case (2) also includes the situation where the $Z'$ decaying into the SM particles has a large displacement vertex outside the detectors.
Finally, Case (3) illustrates the potential usefulness of our kinematic method even for an extremely light $Z'$ that can mediate a long-range interaction. In fact, our results show that it is feasible to probe this case while most collider searches looking for visible particles would not be able to. Of course, such a case would be subject to additional constraints such as an alteration of the Casimir effects, which is beyond the scope of this paper. Nevertheless, the capability of probing the long-range interaction (in general, extremely light particles) at the high energy experiments is an attractive feature of the kinematic method.

On top of the above-mentioned, our kinematic study can cover even more generic scenarios such as
\begin{itemize}
\item[(a)] $Z'$ being heavier than the typically assumed values (i.e., order of $1 \gev$), and \vspace{-0.1cm}
\item[(b)] $Z'$ decayed from a $H^{\pm}$ via a light, non-SM scalar ($h$) in all on-shell process, $H^\pm \to W^\pm h \to W^\pm Z' Z'$ with ${\rm BR} (h \to Z' Z') \sim 1$ \cite{Lee:2013fda}.
\end{itemize}
Along this line, we anticipate that our technique is not
restricted to a particular assignment of spin to the dark force mediators. For concreteness, the detailed analysis in this paper will be performed with $Z'$s having the mass up to $20 \gev$. Based upon the performance, we then address the applicability of the main idea for scalar mediators.

In spite of various interesting scenarios with (invisible) light dark force carriers, relevant collider searches in the top decays are, in general, rather challenging. The reason is that their invisibility causes the associated collider signature (e.g., $t \to b W + Z'$s) to be almost indistinguishable from the dominant SM top decay ($t \to b W$), which results in the poor separation of the signal events from the background ones. 
In order to enhance the relevant signal sensitivity, we adopt the on-shell constrained $M_2$ variables~\cite{Barr:2011xt,Mahbubani:2012kx,Cho:2014naa}, which have been originally proposed in the context of a $(3+1)$ dimensional analogue of the $(2+1)$ dimensional $M_{T2}$ variable~\cite{Lester:1999tx,Cho:2007qv,Burns:2008va,Kim:2009si}. 
The main idea behind the associated search scheme can be summarized as follows. The $M_2$ variables to be utilized are constructed with the assumptions of the dileptonic $t\bar{t}$-like {\it symmetric} event topology shown in the left panel of Figure~\ref{fig:decayTopo}. Therefore, if the signal process stems from a different event topology such as an {\it asymmetric} one, then some contradictory results emerge from the relevant distributions, which enables us to discern the signal and background events more effectively.\footnote{Several studies that attempt to extract useful information from asymmetric event topologies have performed, for example, Ref.~\cite{Agashe:2010tu} in the context of distinguishing dark matter stabilization symmetry and Refs.~\cite{Graesser:2012qy,Cho:2014yma} in the context of supersymmetric top quark partner search. } In fact, such a different event topology arises very naturally in the process of our interest. Due to the smallness of the branching ratio for $t \rightarrow b H^{\pm}$, {\it negligible} is the chance that both tops involve the $Z'$ in their final state. Hence, the dominant signal process is characterized by the mixture between an ordinary and a rare top decay (i.e., $t \bar t \to b W^+ \bar b W^- Z'$), which differs completely from the typical $t\bar{t}$ decays (see the right panel of Figure~\ref{fig:decayTopo}).

The paper is organized as follows. In the next section, we discuss a $Z'$ model in the context of the analysis presented later. In Section~\ref{sec:reviewM2} we briefly review $M_{2}$ variables along with two important kinematic features. We then provide our simulation results and relevant analyses using $M_2$ variables in Section~\ref{sec:simulation}. Section~\ref{sec:conclusion} is reserved for our conclusions.

%%%%%%%%%%%%%%%%%%%%%%%%%%%%%%%%%%%%%%%%%
%%%%%%%%%%%%%%%%%%%%%%%%%%%%%%%%%%%%%%%%%
\section{\label{sec:model} Theoretical Setup}
%%%%%%%%%%%%%%%%%%%%%%%%%%%%%%%%%%%%%%%%%
%%%%%%%%%%%%%%%%%%%%%%%%%%%%%%%%%%%%%%%%%
The specific model we consider is the so-called ``Dark $Z$'' model introduced in Ref.~\cite{Davoudiasl:2012ag}.
The model is based on the Type-I two Higgs doublet model with an additional $U(1)$ gauge boson and a Higgs singlet of the dark sector.
Various physics discussions of the model can be found in Refs.~\cite{Davoudiasl:2012ag,Davoudiasl:2012qa,Lee:2013fda,Davoudiasl:2014mqa,Kong:2014jwa,Davoudiasl:2014kua}.
We closely follow the notations and strategies given in Ref.~\cite{Kong:2014jwa} in our discussions.
Basically, the $Z'$ gauge boson couples to the SM fermions via the mixing between $Z'$ and SM gauge bosons. The relevant interactions between them are described by
\beq
{\cal L}_{\text{dark } Z} = - \left[ \eps e J^\mu_{EM} + \eps_Z (g / \cos\theta_W) J^\mu_{NC} \right] Z'_\mu \label{eq:darkZ}
\eeq
with
\begin{align}
J_\mu^{EM} &= Q_f \bar f \gamma_\mu f \\
J_\mu^{NC} &= (\frac{1}{2}T_{3f} - Q_f \sin^2\theta_W ) \bar f \gamma_\mu f - (\frac{1}{2} T_{3f}) \bar f \gamma_\mu \gamma_5 f
\end{align}
where $\eps$ and $\eps_Z$ are the parametrizations of the effective $\gamma - Z'$ mixing and $Z - Z'$ mixing, respectively.
Here the $J_\mu^{EM}$ ($J_\mu^{NC}$) is nothing but the standard electromagnetic (weak neutral) current.

For simplicity, we take the decoupling limit of the Higgs singlet as in Ref.~\cite{Kong:2014jwa}. The doublet scalars $\Phi_1$ and $\Phi_2$ form two neutral Higgs bosons $h$ and $H$ plus charged Higgs bosons $H^\pm$.
Depending on the choice of parameters, the SM-like Higgs boson can be identified as either the lighter one ($h$) or the heavier one ($H$). Unlike the one identified as the SM-like Higgs, the other neutral scalar couples to the SM particles only through a small mixing between the two doublets. Also, the ratio of the vacuum expectation values denoted by $\tan\beta \equiv v_2 / v_1 \gsim 1$ is required, for which $\Phi_2$ is assumed to couple to the SM fermions while $\Phi_1$ is not.

Depending on the scalar masses in the model, the dominant decay mode of the charged Higgs bosons can be either (i) $H^\pm \to W Z'$ (when the SM-like Higgs boson is the lighter one $h$)~\cite{Davoudiasl:2014mqa}, or (ii) $H^\pm \to W h \to W Z' Z'$ (when the SM-like Higgs boson is the heavier one $H$)~\cite{Lee:2013fda}. One should note that in both cases, the light $Z'$ can be quite elusive especially in collider study as explained earlier. Moreover, the unusual dominant decay channels for the charged Higgs bosons precludes us from applying the typical bounds so that very light $H^{\pm}$'s are allowed~\cite{Lee:2013fda}.
%%%%%%%%% FIGURE %%%%%%%%
\begin{figure}[tb]
\begin{center}
\includegraphics[width=10.2cm]{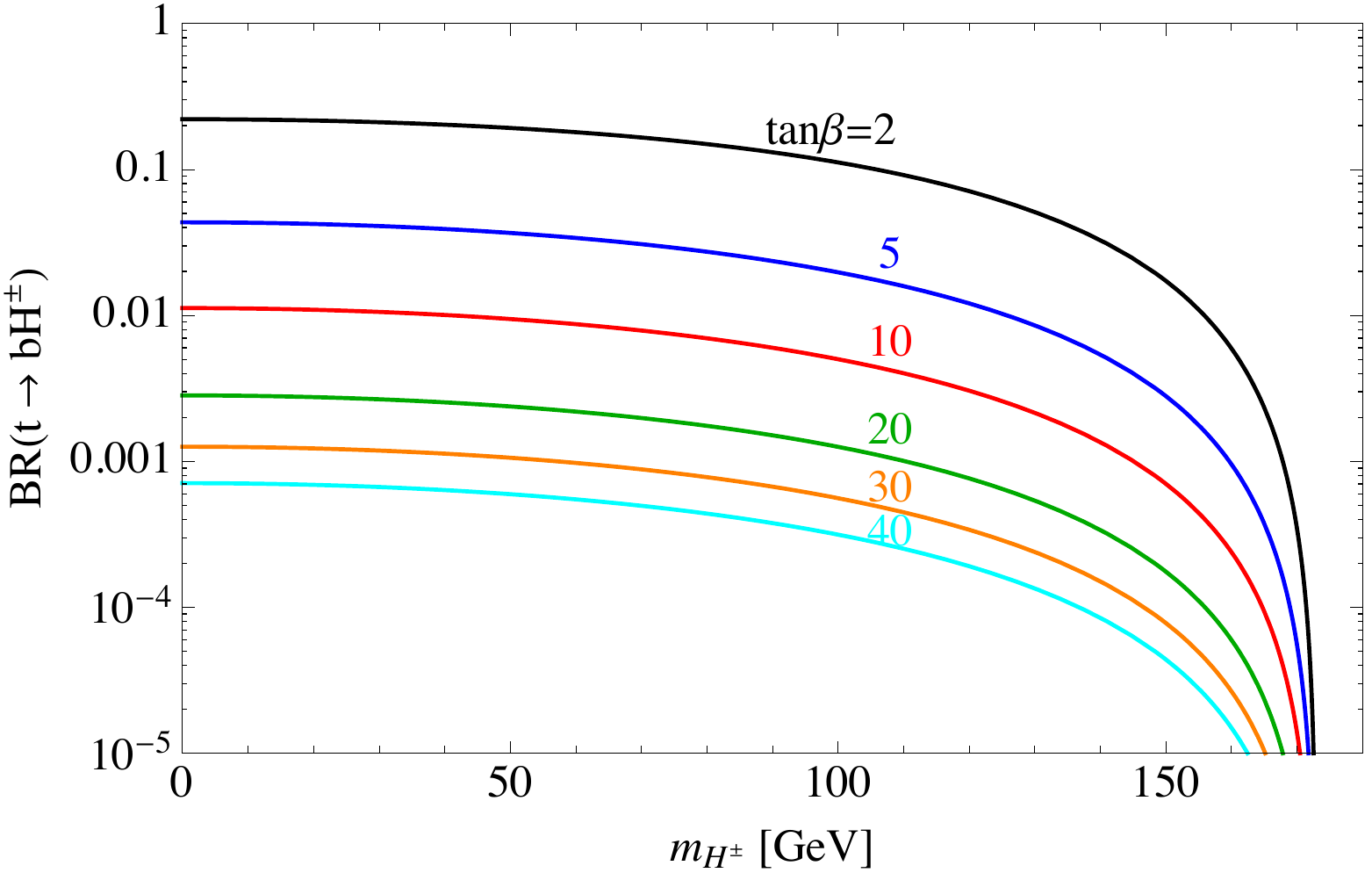}
\end{center}
\caption{Branching ratio of $t \to b H^{\pm}$ for various $\tan\beta$ values. The $H^\pm$ decays dominantly into $W + Z'$s. The $\br(t \to b H^{\pm})$  decreases with $m_{H^\pm}$ and $\tan\beta$.
}
\label{fig:BR}
\end{figure}
%%%%%%%%%%%%%%%%%%%%%%
%
Neglecting $m_b/m_t$ and higher order corrections, we have the decay widths for $t\rightarrow b W$ and $t\rightarrow b H^{\pm}$~\cite{Kong:2014jwa} as follows:
\bea
\Gamma_{t\rightarrow bW}&=&C\left(1-\frac{m_W^2}{m_t^2} \right)^2\left(1+\frac{2m_W^2}{m_t^2} \right) \\
\Gamma_{t\rightarrow bH^{\pm}}&=&C\left(1-\frac{m_{H^\pm}^2}{m_t^2} \right)^2\frac{1}{\tan^2\beta}
\eea
where
\bea
C\equiv\frac{\sqrt{2}G_F |V_{tb}|^2}{16\pi}m_t^3
\eea
with $\tan\beta \gsim 1$ in the dark force model under consideration.
With the assumption that $t\rightarrow bW$ is the dominant decay mode of the top, the branching fraction of $t\rightarrow bH^{\pm}$ for the on-shell decay is
\beq
\br(t \to b H^{\pm}) \approx \left( \frac{m_t^2 - m_{H^\pm}^2}{m_t^2 - m_W^2} \right)^2 \frac{1 / \tan^2\beta}{1 + 2 m_W^2 / m_t^2}. \label{eq:tTobCH}
\eeq
In Figure~\ref{fig:BR}, we show its functional dependence over the charged Higgs mass for several $\tan\beta$ values.
As we can see, $\br(t \to bH^{\pm}) \gsim O(10^{-3})$, which is roughly what we want to explore, can be obtained in a wide range of parameters.
For the large $\tan\beta \gsim 30$, however, the Drell-Yan $H^+H^-$ process has a comparable production cross section \cite{Davoudiasl:2014mqa,Kong:2014jwa}.

Due to the smallness of $\br(t\rightarrow bH^{\pm})$, the signal process is defined as a rare decay of the top quark:
\bea
t \to b H^{\pm} \to b W + Z'\text{s} \to b \ell \nu + Z'\text{s} \label{eq:sigProcess}
\eea
with $\ell = e$, $\mu$, while $Z'$ appears as an extra missing energy in the final state. 
The cross section of our signal in the dileptonic $t \bar t$ channel is given by
\begin{align}
&\sigma(p p \to t \bar t \to b W^\pm \, \bar b H^\mp \to b \ell^+ \bar b \ell^- \nu \bar\nu +Z'\text{s}) \nn \\
&\simeq 2 X [\br(W \to \ell \nu)]^2 \, \sigma_{t \bar t} \\
&\simeq (87 \text{ pb}) \, X
\end{align}
where the 14 TeV LHC $t \bar t$ production cross section $\sigma_{t \bar t} \simeq 953.6$ pb (see Section~\ref{sec:simulation}) was used in the last line. Here $X$ is the parametrization for the branching fraction of the top decay into $Z'$:
\bea
X &\equiv& \br(t \to b W + Z'\text{s}) \label{eq:Xdef} \\
&=& \br(t \to b H^{\pm}) \cdot \br (H^\pm \to W + Z'\text{s}).
\eea
We remark that for the subsequent decay of $H^{\pm}$, a sizable decay branching ratio $\br (H^\pm \to W + Z'\text{s}) \simeq 0.5 - 1$ can be obtained in a wide range of parameter space.
We refer to Ref.~\cite{Kong:2014jwa} and references therein for further details.

%%%%%%%%%%%%%%%%%%%%%%%%%%%%%%%%%%%%%%%%%
%%%%%%%%%%%%%%%%%%%%%%%%%%%%%%%%%%%%%%%%%
\section{\label{sec:reviewM2} \boldmath General strategy with $M_2$ variables}
%%%%%%%%%%%%%%%%%%%%%%%%%%%%%%%%%%%%%%%%%
%%%%%%%%%%%%%%%%%%%%%%%%%%%%%%%%%%%%%%%%%
We now provide a brief review on the (on-shell constrained) $M_2$ variables that are employed for the analysis in the next section. Particular attention is paid upon a couple of kinematic features which will be the basic ingredients of our strategy for discriminating signal events from background ones. 
Since the SM dileptonic $t\bar{t}$ is the most challenging background to our signal process as briefly mentioned in the introduction, we describe the $M_2$ variables by taking $t\bar{t}$ itself as a concrete example:
\bea
t_i \rightarrow b_iW_i \rightarrow b_i \ell_i \nu_i \;\;(i=1,\;2).
\eea
Here $i$ is simply the index indicating the associated decay side, {\it not} implying different masses of the corresponding particles. The full decay topology is also sketched in the left panel of Figure~\ref{fig:decayTopo}.

The $M_2$ variables~\cite{Barr:2011xt,Mahbubani:2012kx,Cho:2014naa} have been recently proposed as a $(3+1)$ dimensional analogue of the $(2+1)$ dimensional $M_{T2}$ variable~\cite{Lester:1999tx,Cho:2007qv,Burns:2008va,Kim:2009si}. More specifically, for each event, they are defined as 
\begin{itemize}
\item $M_2$: a minimization of the maximum of the two invariant masses in both decay chains under the $\met$ constraint
\end{itemize}
and some (optional) equal mass constraints that will be explained shortly. Due to the similarity between $M_{T2}$ and $M_2$ variables, one can define $M_2$ for three different subsystems, namely, ($b$), ($\ell$), and ($b\ell$) subsystems~\cite{Burns:2008va}, which were originally named after the visible particles associated with the subsystem under consideration. According to Ref.~\cite{Cho:2014naa}, for each subsystem, the particle whose mass is minimized over is denoted as ``parent'' ($P_i$), while the particle whose mass is hypothesized is denoted as ``child''. The remaining particle is denoted as ``relative'' ($R_i$). For instance, in the ($b$) subsystem, we have $t = $ parent, $W = $ child, and $\nu = $ relative. In the ($b\ell$) subsystem, we have $t =$ parent, $\nu =$ child, and $W = $ relative.

%%%%%%%%% FIGURE %%%%%%%%
\begin{figure}[t]
\centering
\includegraphics[width=7.2cm]{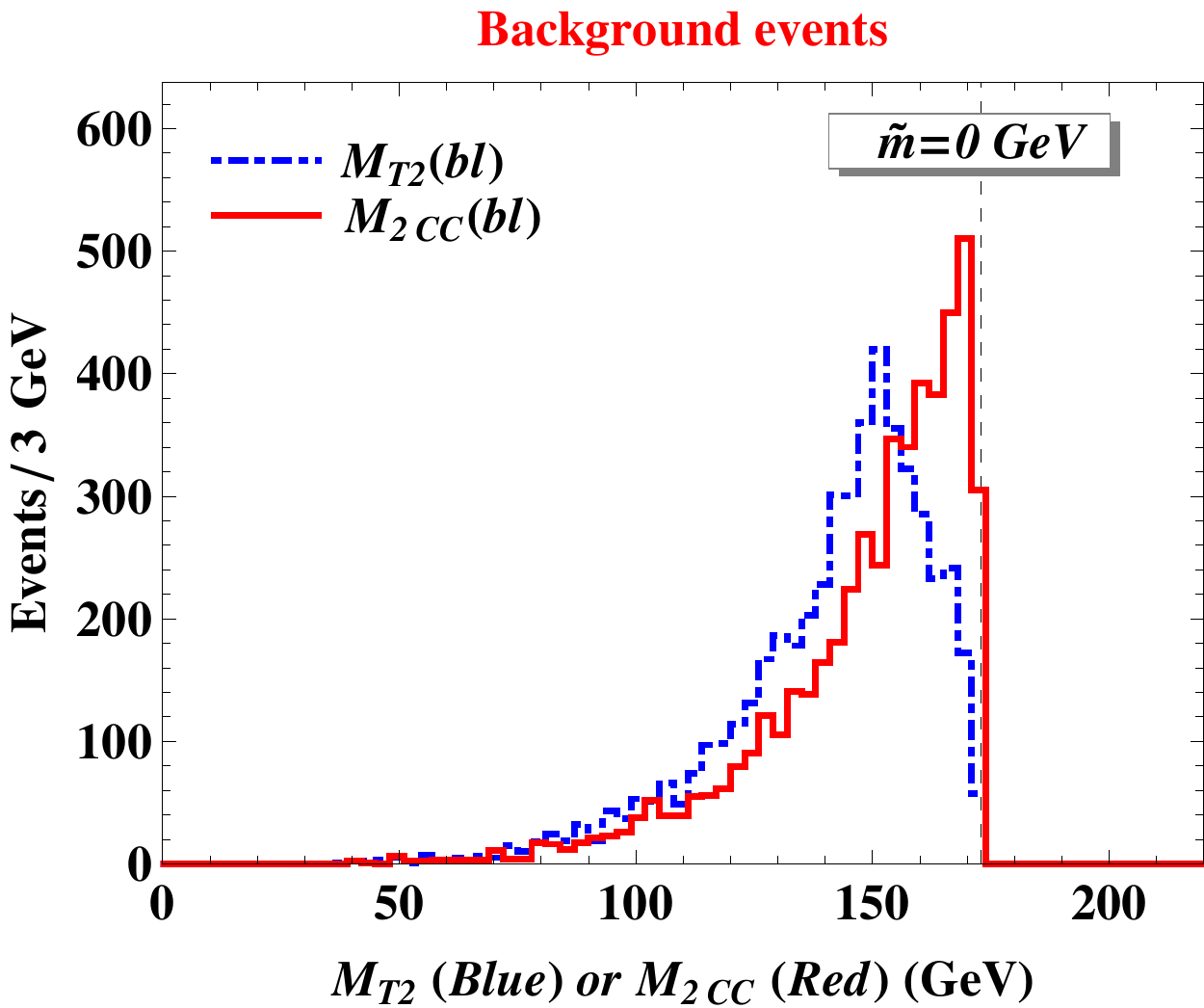} \quad
\includegraphics[width=7.2cm]{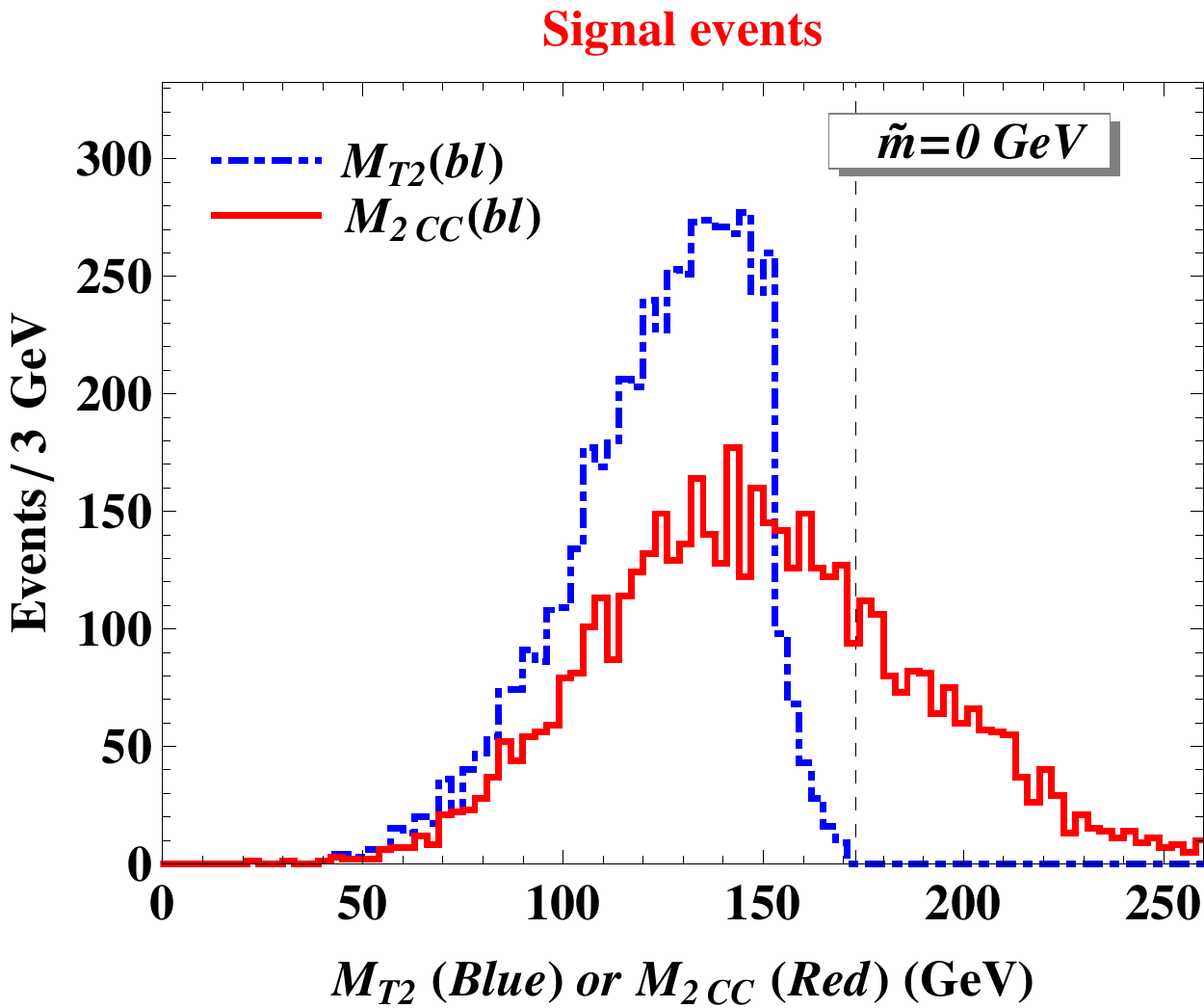}
\caption{\label{fig:partonlevel} Comparisons of $M_{T2}$ (blue dashed) and $M_{2CC}$ (red solid) distributions with 10,000 parton level events for the ($b\ell$) subsystem. The left panel shows the case of $t\bar{t}$ system, whereas the right panel shows the case of our signal process. For the distributions in the right panel, 130 GeV and 1 GeV are chosen for the masses of $H^{\pm}$ and $Z'$. The test mass for the child particle is set to be $\tilde m = 0$ GeV under the (naive) assumption that the child particle is the neutrino. The vertical dashed lines denote the expected endpoints of $M_{T2}$ distributions for the chosen mass spectrum. }
\end{figure} 
%%%%%%%%%%%%%%%%%%%%%%

Computation of $M_2$ values typically involves a numerical minimization as $M_{T2}$ does, and a calculation package is available in Ref.~\cite{Cho:2015laa}. One of the distinctive features of $M_2$ from $M_{T2}$ is that the associated numerical procedure yields the {\it ansatz} for the full momenta of the two invisible particles due to the $(3+1)$ dimensionality of $M_2$ variables. Unlike the $M_{T2}$, this actually enables us to reconstruct the mass of the relative particles together with the full momenta of visible particles belonging to the same decay chain.   

One should note that, like $M_{T2}$, the mass of the invisible particle is not known {\it a priori} in the general situation so that a ``test'' mass (henceforth, denoted by $\tilde{m}$) should be introduced when the $M_2$ value is evaluated, and as a result $M_2$ is also given by a function over $\tilde{m}$. Usually, a single type of the test mass is assumed, and thus the masses of child particles are trivially the same. On the contrary, we have two non-trivial options of imposing equal mass conditions on the parent and relative particles as mentioned earlier:
\bea
M_{P_1}&=&M_{P_2} \label{eq:equalP}\\
M_{R_1}&=&M_{R_2} \label{eq:equalR}
\eea
where $M_{P(R)_{1(2)}}$ is the parent (relative) mass in decay side 1 (2). Since these two additional constraints define different $M_2$ variables, for the identification purpose we add a subscript ``C'' or ``X'' that indicates whether Eq.~\eqref{eq:equalP} is demanded or not (respectively), followed by a second ``C'' or ``X'' that does whether Eq.~\eqref{eq:equalR} is demanded or not. For example, $M_{2XC}(\ell)$ implies that it is built in the ($\ell$) subsystem with only the masses of relative particles (here $\nu_i$) are forced to be the same. For more formal definitions, we refer to Ref.~\cite{Cho:2014naa}.

A couple of kinematic properties of $M_2$ variables should be highlighted, which are crucial ingredients for our collider study. Let us first suppose that the actual physics is consistent with the model hypotheses that the relevant $M_2$ variables take. In such a case, it has been explicitly proven that the following hierarchy among $M_{T2}$ and various $M_2$ variables holds for each event~\cite{Cho:2014naa}.
\bea
M_{T2} =M_{2XX}=M_{2CX} \leq M_{2XC} \leq M_{2CC} \label{eq:hier1}
\eea
where inequalities become equalities at the associated kinematic endpoint:
\bea
M_{T2}^{\max} =M_{2XX}^{\max}=M_{2CX}^{\max} = M_{2XC}^{\max} = M_{2CC}^{\max}. \label{eq:hier2}
\eea
Heuristically, this property can be understood that there is a better chance for the $M_{2CC}$ to find the true solution, compared to the $M_{T2}$ because the same constraints as the actual event can be imposed. The physical implication encoded in this mathematical relationship is that more events get populated near the relevant kinematic endpoints as more constraints are applied. In other words, the endpoint becomes sharper but {\it no} events are migrated beyond it. 

Obviously, the above-given observation is relevant to the $t\bar{t}$ events since we have constructed the $M_2$ variables of interest upon the assumption of the dileptonic decay topology of $t\bar{t}$. A Monte Carlo simulation demonstrated in the left panel of Figure~\ref{fig:partonlevel} confirms such an expectation. Here 10,000 events were generated at the parton level with the mass spectrum being exactly the same as the $t\bar{t}$ system, i.e., $(m_t,\;m_W,\;m_\nu)=(173,\;80,\;0)$ GeV and the test mass fixed to be $\tilde m = 0$ GeV matching to the true neutrino mass. We clearly see that the endpoint structure of the ($b\ell$) subsystem gets more sharpened in the $M_{2CC}$ distribution (red solid), but no events exceed the corresponding $M_{T2}$ endpoint (vertical dashed), which is the same as the parent mass ($173 \gev$) as the test mass is the same as the neutrino mass.

On the other hand, once the model assumptions differ from the actual physics, e.g., decay processes of mother particles via a three-body decay or via different intermediate states, the consequence is more dramatic. More specifically, for the ($b\ell$) subsystem the relevant model assumptions can be rephrased as follows:
\begin{itemize}
\item there exists an intermediate resonance in each decay leg, and 
\item the two intermediate states have the same mass.
\end{itemize}
If the actual event topologies at hand are inconsistent with either of the above two, $M_{2XC}$ and $M_{2CC}$ variables lose their physical meaning so that it is possible for some events to give rise to $M_2$ values {\it beyond} the expected kinematic endpoint. In other words, the relation~\eqref{eq:hier1} is still true, whereas the full equality~\eqref{eq:hier2} is no longer guaranteed.  
Our $Z'$ signal can be classified to this case because one decay side is proceeded via an on-shell $W$ gauge boson ($1:~ t \to b W \to b \ell \nu$), while the other is proceeded via $H^\pm$ boson ($2:~ t \to b H^{\pm} \to b \ell \nu + Z'$) as also shown in the right panel of Figure~\ref{fig:decayTopo}. The right panel of Figure~\ref{fig:partonlevel} clearly exhibits that the $M_{2CC}$ variable (red solid) can ``violate'' the corresponding $M_{T2}$ endpoint (vertical dashed) for the case where the true event topology differs from the associated model assumptions. Again, 10,000 events were generated as in the $t\bar{t}$ case. The identical mass spectrum to that for the representative benchmark point (BP) in the next section is chosen; the mass of the charged Higgs is set to be 130 GeV while the mass of $Z'$ is set to be 1 GeV. The test mass is set to be $\tilde m = 0$ GeV as before.

The above two observations actually suggest a clever strategy that we pursue in the detailed analysis explicated in the following section. As mentioned earlier, the $M_2$ variables to be employed is devised targeting on the dileptonic $t\bar{t}$ decay topology. Among them we choose the $M_{2CC}$ to maximally constrain the system of interest. Since the $t\bar{t}$ background events originate from the same decay topology, most of the events are confined to the regime below its $M_{T2}$ endpoint. In contrast, the signal events are in contradiction to the model hypotheses that the $M_{2CC}$ variable bears, and as a consequence, many of them are anticipated to exceed the $M_{T2}$ endpoint of the $t\bar{t}$ system, which leads to an enhancement of $S/B$ along with an optimal choice of $M_{2CC}$ cuts.

%%%%%%%%%%%%%%%%%%%%%%%%%%%%%%%%%%%%%%%%%
%%%%%%%%%%%%%%%%%%%%%%%%%%%%%%%%%%%%%%%%%
\section{\label{sec:simulation} Simulation results and discussions}
%%%%%%%%%%%%%%%%%%%%%%%%%%%%%%%%%%%%%%%%%
%%%%%%%%%%%%%%%%%%%%%%%%%%%%%%%%%%%%%%%%%

Here, we discuss the discovery potential of the light $Z'$ using $M_2$ variables reviewed in the previous section together with Monte Carlo simulations. For the purpose of a more realistic study we take cuts and detector resolutions into consideration. The parton level event generation is done by \texttt{MadGraph\_aMC@NLO}~\cite{Alwall:2014hca} where parton distributions inside protons are evaluated by the default \texttt{NNPDF23}~\cite{Ball:2012cx}, and the relevant output is fed to \texttt{Pythia6.4}~\cite{Sjostrand:2006za} and \texttt{Delphes3}~\cite{deFavereau:2013fsa} in order. 
All the simulation is conducted with a $pp$ collider of $\sqrt{s}=14$ TeV at the leading order. The cross section for $t\bar{t}$, $\sigma_{t\bar{t}}$ is rescaled to the predicted one which for $pp$ collisions at the given center-of-mass energy is 953.6 pb for a top quark mass of 173 GeV, which is calculated at next-to-next-to-leading order (NNLO) in QCD including resummation of next-to-next-to-leading logarithmic (NNLL) soft gluon terms with \texttt{Top++2.0}~\cite{Cacciari:2011hy,Baernreuther:2012ws,Czakon:2012zr,Czakon:2012pz,Czakon:2013goa,Czakon:2011xx}. The signal production cross section $\sigma_{Z'}$ is computed with $X = \br(t\rightarrow b W + Z'\text{s})$ of Eq.~\eqref{eq:Xdef}. Again, since $X$ is expected to be small, the chance of having both top quarks decayed via the process in Eq.~\eqref{eq:sigProcess} is negligible, and thus we have $\sigma_{Z'}\cong2 X \sigma_{t \bar t}$.

Provided with the final state defined by the signal process of interest, i.e., $b\bar{b}\ell^+\ell^-+\met$, there are several SM backgrounds to be considered. It turns out that among them $t\bar{t}$ is the dominant irreducible background with the aid of the selection criteria that will be explained shortly.\footnote{Single top production via $tW$ becomes the sub-leading background to our signal process after applying the selection cuts. We find that its cross section after the cuts in (\ref{eq:cut1}) through~(\ref{eq:cut4}) is smaller than the corresponding cross section for $t\bar{t}$ by a factor of 40. Therefore, its contribution does not affect $S/\sqrt{B}$ for discovery in conjunction with the luminosities of our interest. So, we ignore its contribution for the later analysis.  } In order to avoid any possible unwanted endpoint violation from backgrounds other than $t\bar{t}$, we employ a rather hard event selection scheme similar to that in Ref.~\cite{ATLAS} for the top mass measurement in dileptonic top quark pair decays. The key criteria are enumerated below with slight modifications:
\bea
&&\hspace{-1cm}\textnormal{$\bullet$ $N_\ell = 2$ with opposite signs, $p_T^e>25$ GeV, and $p_T^{\mu}>20$ GeV,}\label{eq:cut1}\\ 
&&\hspace{-1cm}\textnormal{$\bullet$ $\met>60$ GeV for the $ee/\mu\mu$ channels and $H_T >130$ GeV for the $e\mu$ channel,}\\ 
&&\hspace{-1cm}\textnormal{$\bullet$ $m_{ee/\mu\mu}>15$ GeV and $|m_{ee/\mu\mu}-m_Z|>10$ GeV,}\\ 
&&\hspace{-1cm}\textnormal{$\bullet$ $N_j \geq 2$ while $N_b =2$, $p_T^j>25$ GeV, and $|\eta^j|<2.5$}\label{eq:cut4}
\eea
where $N_{\ell}$ and $N_{j(b)}$ denote the number of selected leptons and jets ($b$-tagged jets), respectively, and $H_T$ is defined as $\sum_{i=\ell,j}p_T^i$. Jets are built by the anti-$k_t$ algorithm~\cite{Cacciari:2008gp} together with a radius parameter $R=0.4$, and the $b$-tagging efficiency is taken to be 70\%, while the light quark jets are mis-tagged by a rate of 1/130. The $M_{2CC}$ cuts will be applied for the signal and background events passing all the selection criteria given above. The cross section for $t\bar{t}$ after them is estimated to be 2.99 pb, for which the relevant selection efficiency is close to that in Ref.~\cite{ATLAS}.  

%%%%%%%%%%%%%%%% FIGURE %%%%%%%%%%%%%%%
\begin{figure}[t]
\centering
\includegraphics[width=6.8cm]{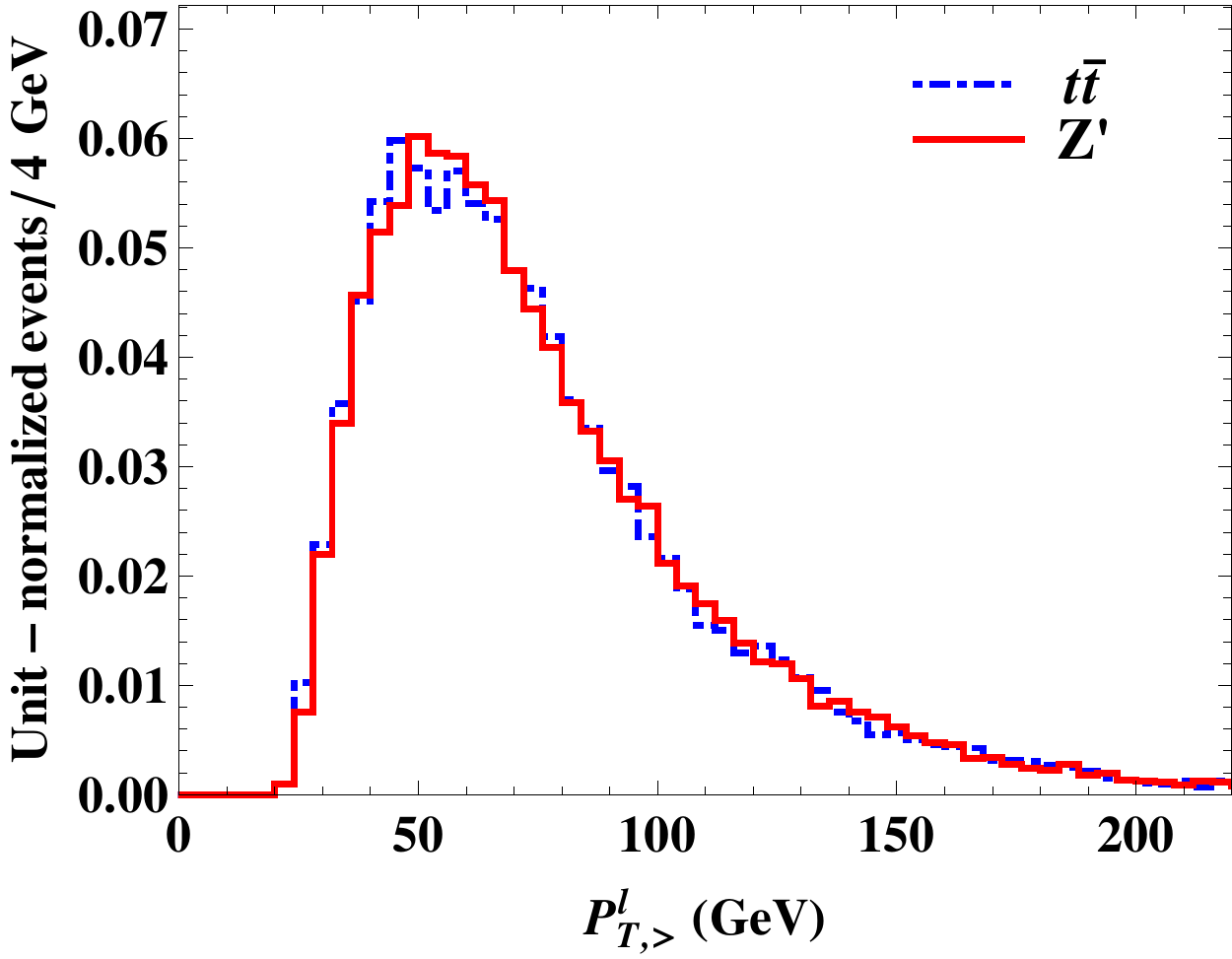}
\includegraphics[width=6.8cm]{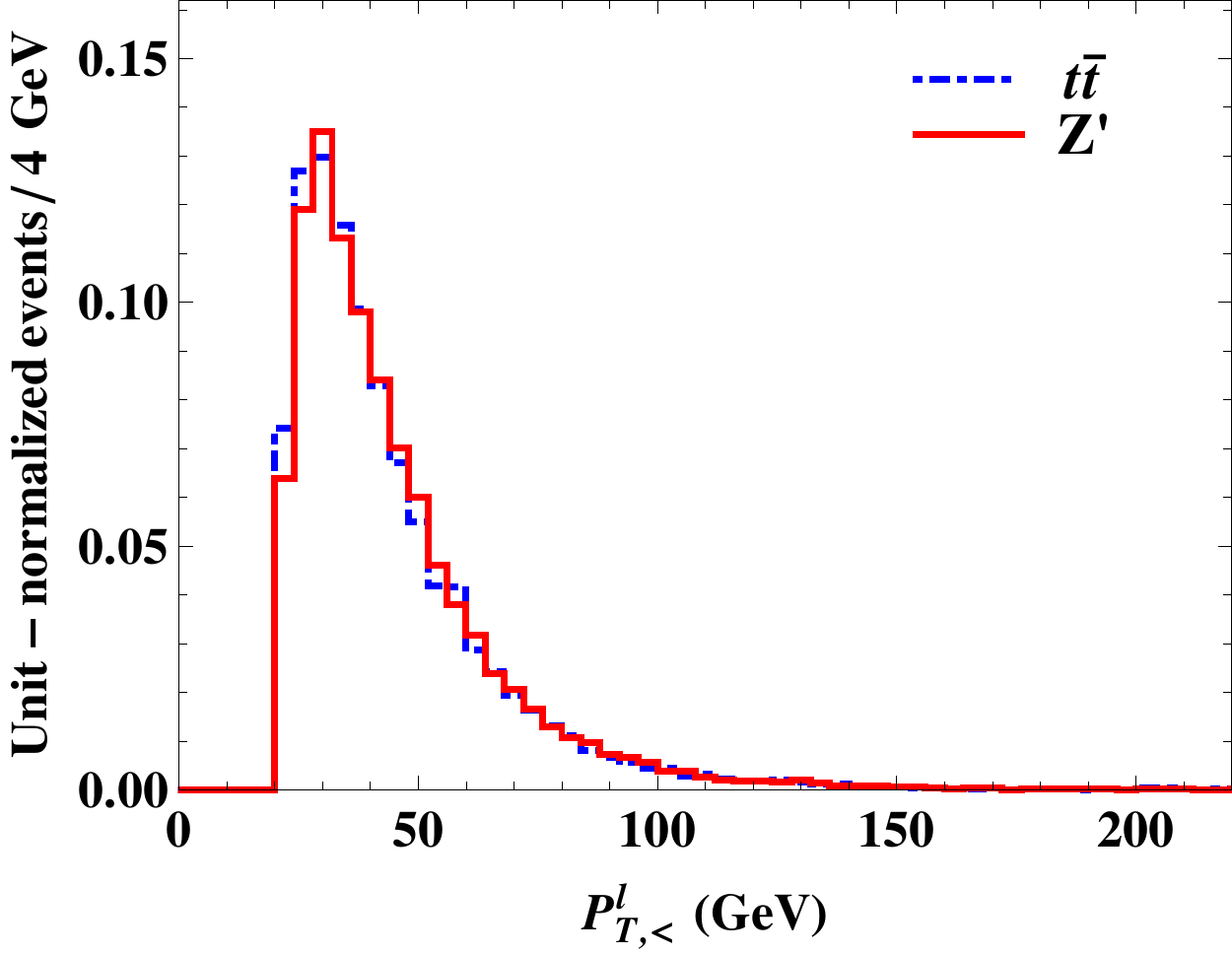}
\includegraphics[width=6.8cm]{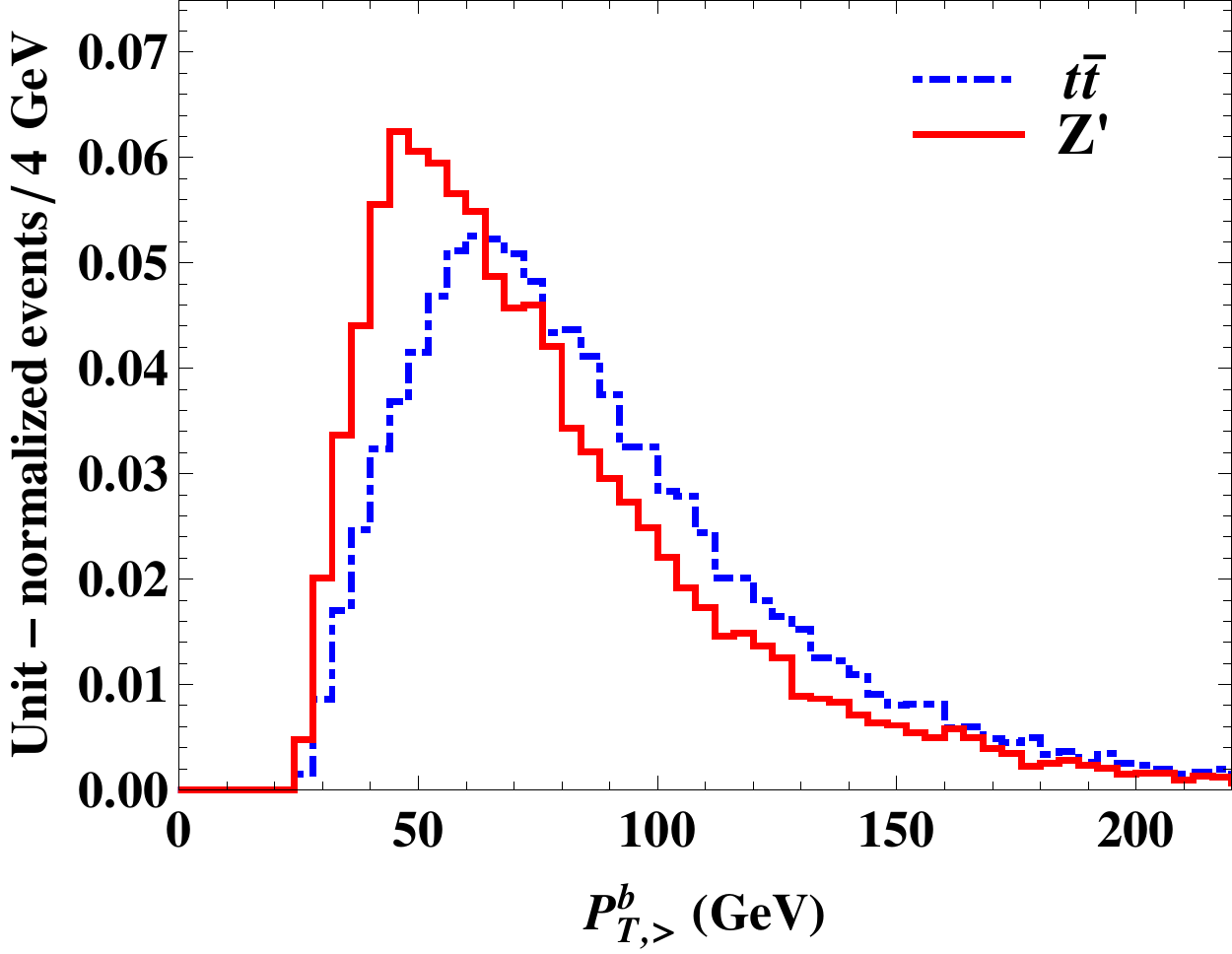}
\includegraphics[width=6.8cm]{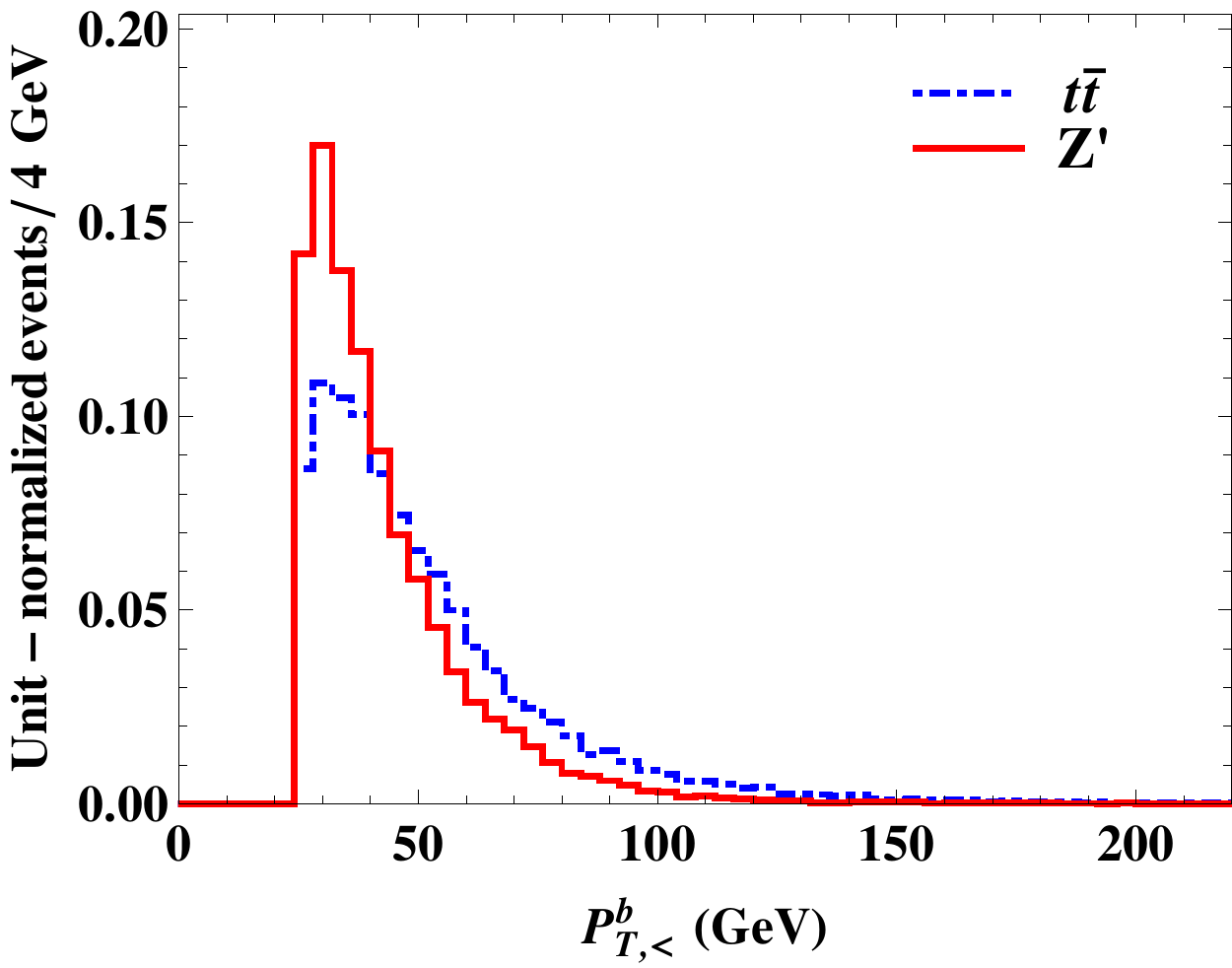}
\includegraphics[width=6.8cm]{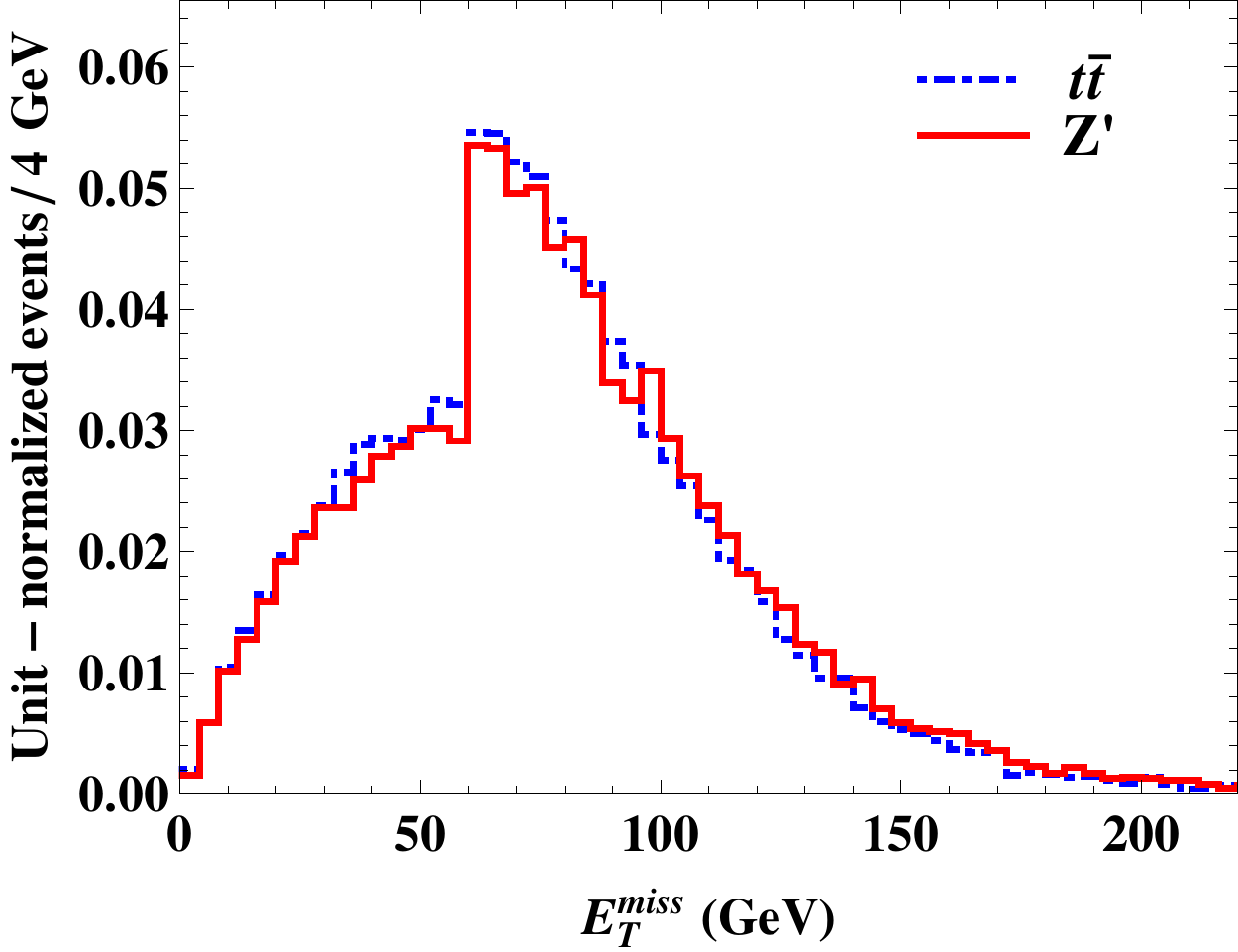}
\includegraphics[width=6.8cm]{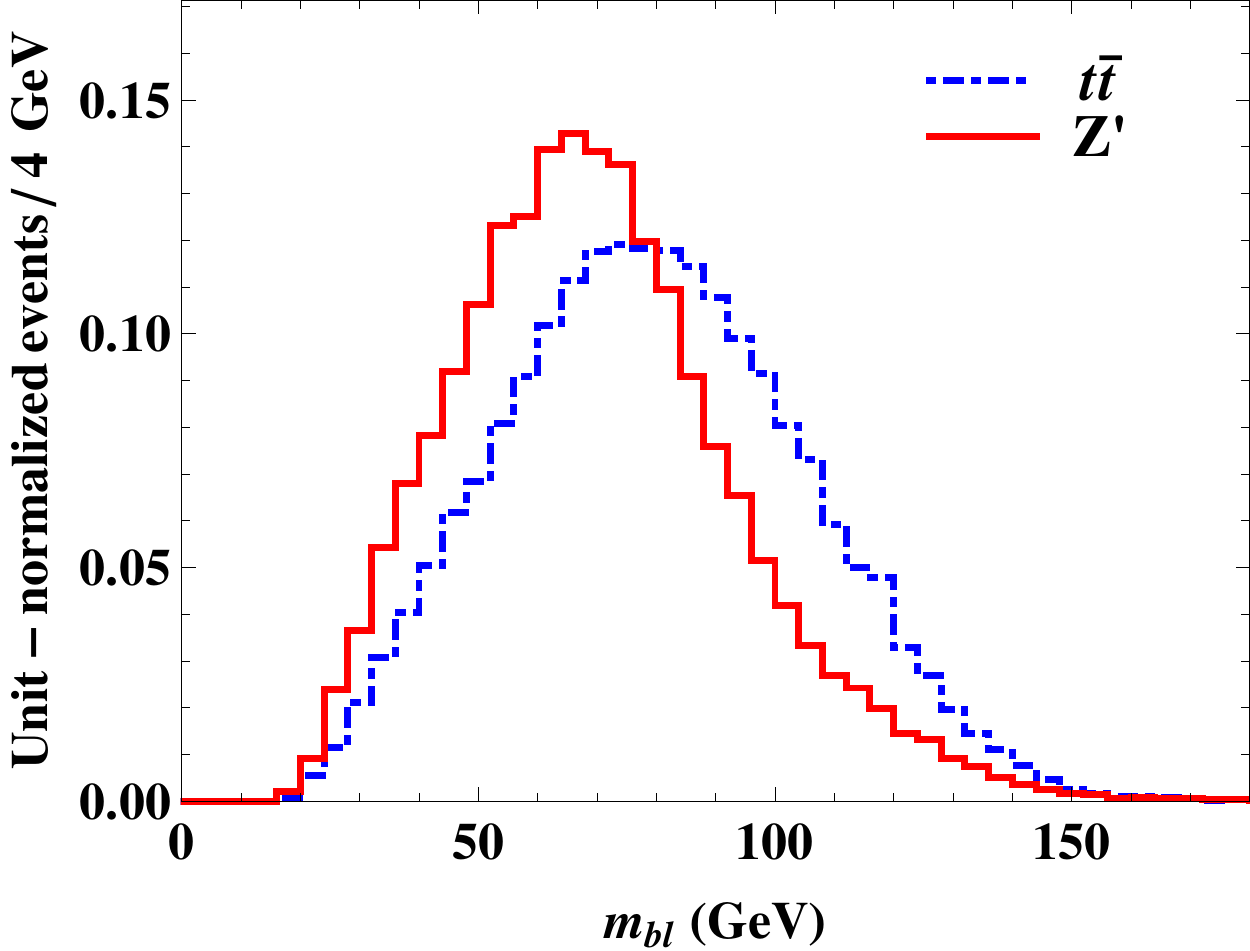}
\caption{\label{fig:stdvar} Comparisons of signal and background behaviors in the harder (top left panel) and the softer (top right panel) $p_T^{\ell}$ distributions, the harder (middle left panel) and the softer (middle right panel) $p_T^{b}$ distributions, the $\met$ distribution (bottom left panel), and the $m_{b\ell}$ invariant mass distribution (bottom right panel). For the invariant mass variable, the two smallest values are taken out of four possible combinations. All the plots are produced with the detector-level events passing all selection cuts enumerated in (\ref{eq:cut1}) through (\ref{eq:cut4}).  }
\end{figure}
%%%%%%%%%%%%%%%%%%%%%%%%%%%%%%%%%%%%%%%

Note again that our study is motivated by the light dark force carriers. In this sense, we demonstrate the detailed performance of our technique later on with a benchmark point (BP1) in Table~\ref{tab:discovery}, in which $Z'$ is nearly massless while the $H^{\pm}$ mass is in-between the top quark and $W$ gauge boson masses. We then perform the same analysis for the other benchmark points, in particular, to examine the sensitivity of the proposed method to the two mass parameters $m_{H^{\pm}}$ and $m_{Z'}$. The possible effect in replacing a vector boson $Z'$ by a non-SM, light scalar $h$ upon our technique will be briefly discussed as well.  

Prior to the application of $M_2$ variables, we first show that conventional variables such as $p_T^{b,\ell}$ and $\met$ would be unsuccessful in discriminating the signal events from the background ones. The top (middle) panels in Figure~\ref{fig:stdvar} demonstrate the lepton (bottom jet) transverse momentum distributions for signal and background events. For more careful comparison, they are decomposed into the harder $(p_{T,>})$ and the softer $(p_{T,<})$ transverse momenta. Speaking of the leptons, the signal and background distributions are almost identical to each other mainly because the leptons are emitted from $W$ gauge boson in both cases, and as a result, the hardness of leptons is not distinctive. On the contrary, the bottom transverse momentum for the signal is typically a little softer than that for the background. The reason is that the mass gap between the top quark and the charged Higgs is smaller than that between the top quark and the $W$ gauge boson so that the $b$-jet in the signal process tends to come out with a smaller momentum. When it comes to $\met$, the overall similarity in $p_T^{b,\ell}$ ensembles is unable to make a discernible difference between the signal and background $\met$ distributions (see the bottom left panel of Figure~\ref{fig:stdvar}). Considering the typical schemes of retaining only the events beyond given $p_T^{b,\ell}$ or $\met$ cuts, therefore, we find that they are not good discriminators. 

One could also consider the invariant mass variable formed by a $b$-jet and a lepton partly because the existence of an extra invisible particle in the final state would give rise to some distinctive feature in the corresponding distribution. One well-defined property is the kinematic endpoint, and as a matter of fact, the analytic formula for the decay leg involving $Z'$ is readily available~\cite{Agashe:2010gt,Cho:2012er}:
\bea
\left(m_{b\ell}^{\max}\right)^2 = \frac{2(m_t^2-m_{H^{\pm}}^2)m_W^2}{m_{H^{\pm}}^2+m_W^2-m_{Z'}^2-\lambda^{1/2}(m_{H^{\pm}}^2,m_W^2,m_{Z'}^2)} \label{eq:ep}
\eea
with the kinematic triangular function being defined as  
\bea
\lambda(x,y,z)\equiv x^2+y^2+z^2-2(xy+yz+zx)
\eea
where lepton and bottom quark are assumed massless for simplicity. One can actually prove that the corresponding endpoint for the ordinary top decay chain is larger than Eq.~(\ref{eq:ep}) for any pairs of $(m_{H^{\pm}}, m_{Z'})$ so that signal events are completely buried in the background $m_{b\ell}$ distribution. One would try a shape analysis because the decay chain involving $Z'$ develops a cusp structure in the middle of the invariant mass distribution~\cite{Agashe:2010gt,Cho:2012er}. Considering realistic effects such as cuts, combinatorics, relatively small signal cross section, and so on, however, such an option is not beneficial, either (see also the bottom right panel of Figure~\ref{fig:stdvar}).  

%%%%%%%%% FIGURE %%%%%%%%
\begin{figure}[H]
\centering 
\includegraphics[width=6.3cm]{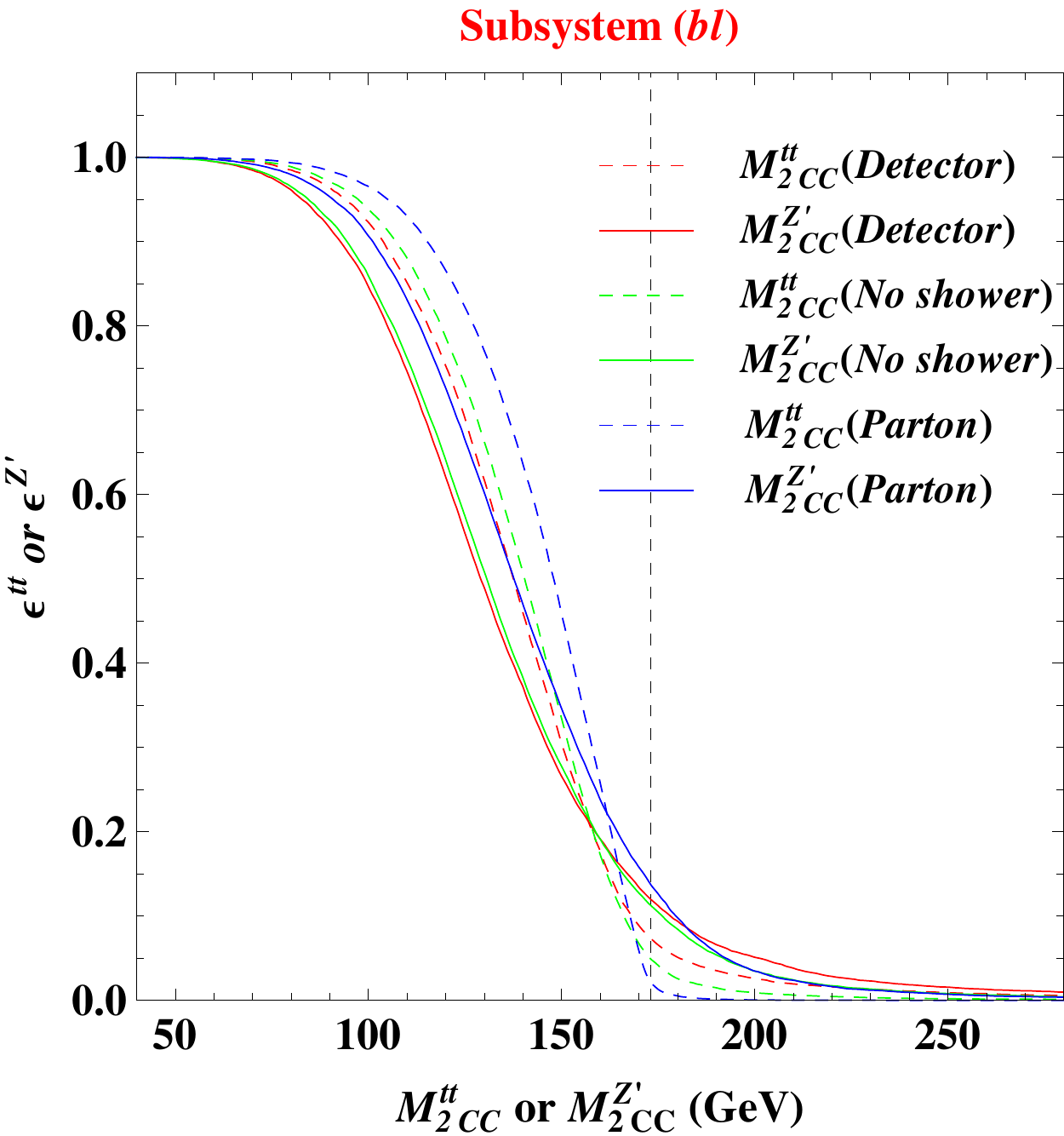}~~~
\includegraphics[width=6.3cm]{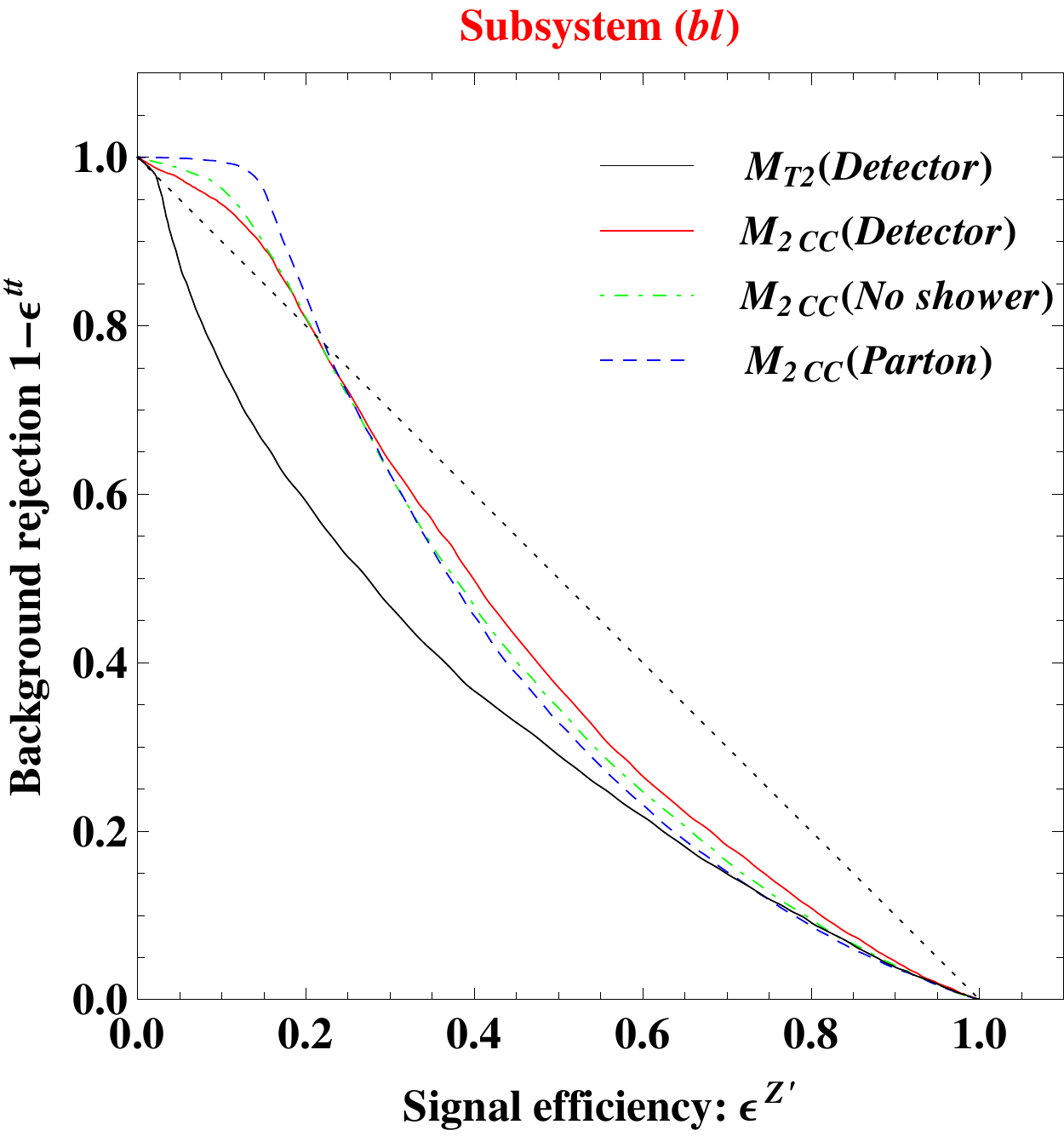}\\
\includegraphics[width=6.3cm]{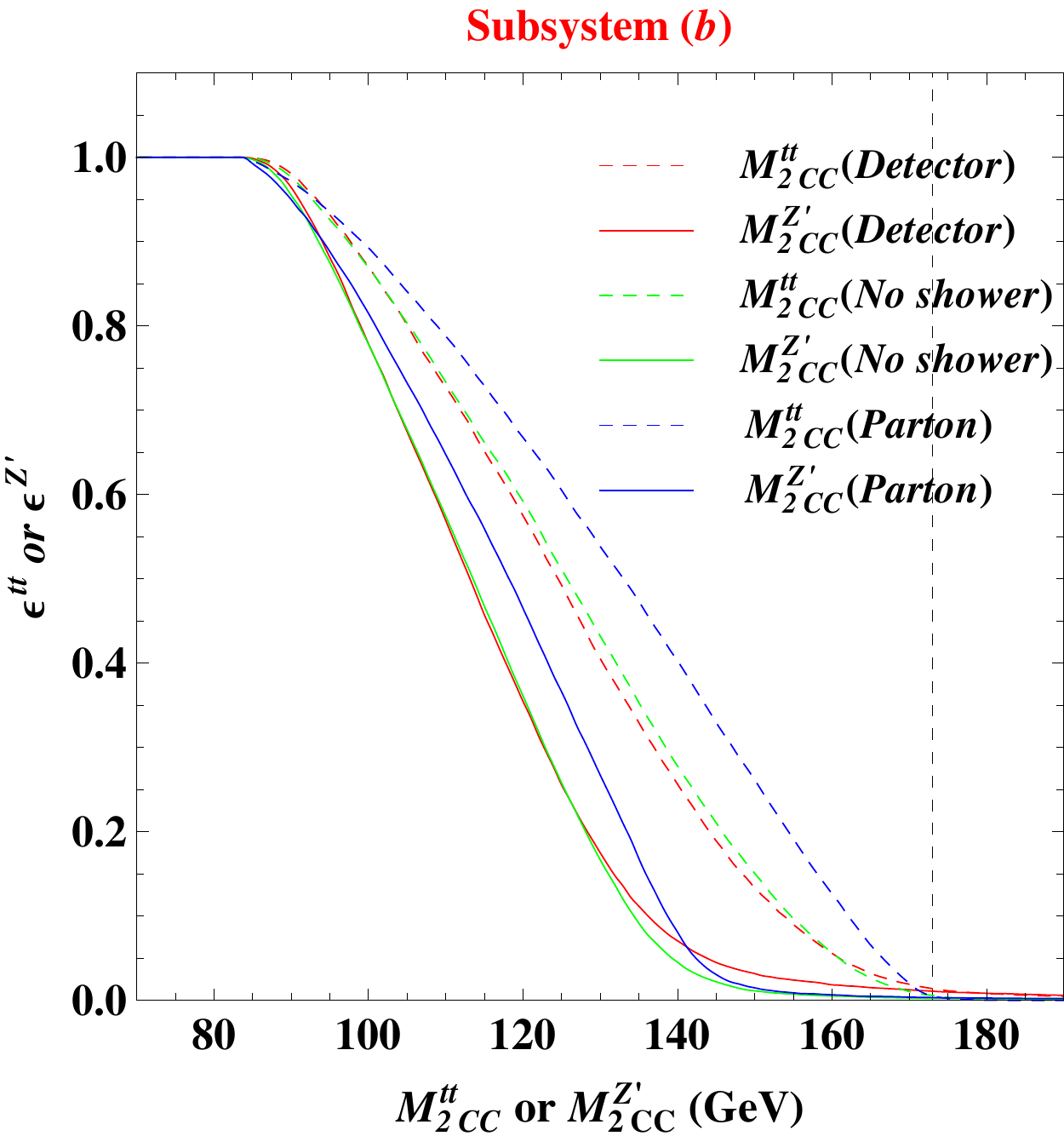}~~~
\includegraphics[width=6.3cm]{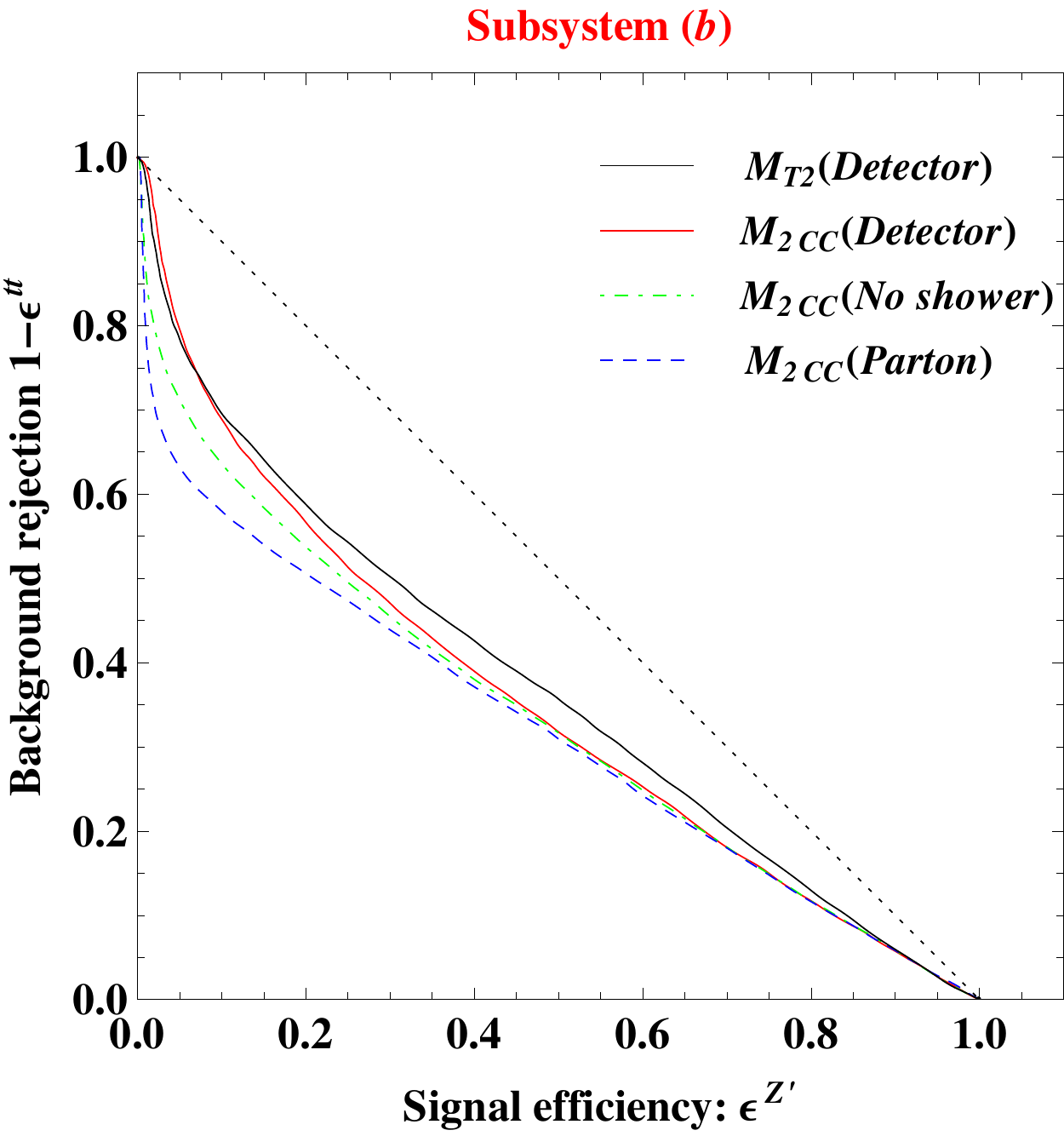}\\
\includegraphics[width=6.3cm]{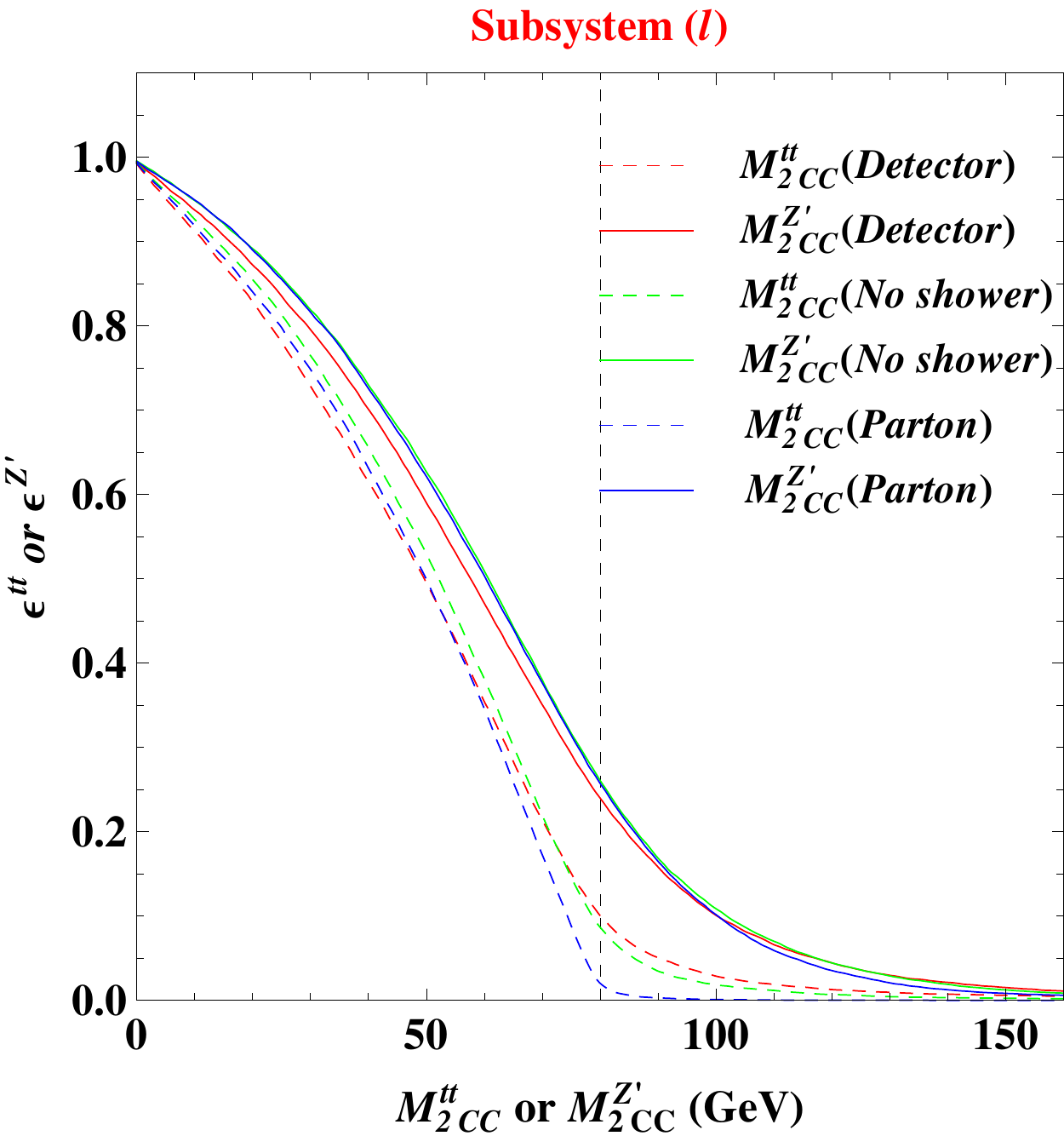}~~~
\includegraphics[width=6.3cm]{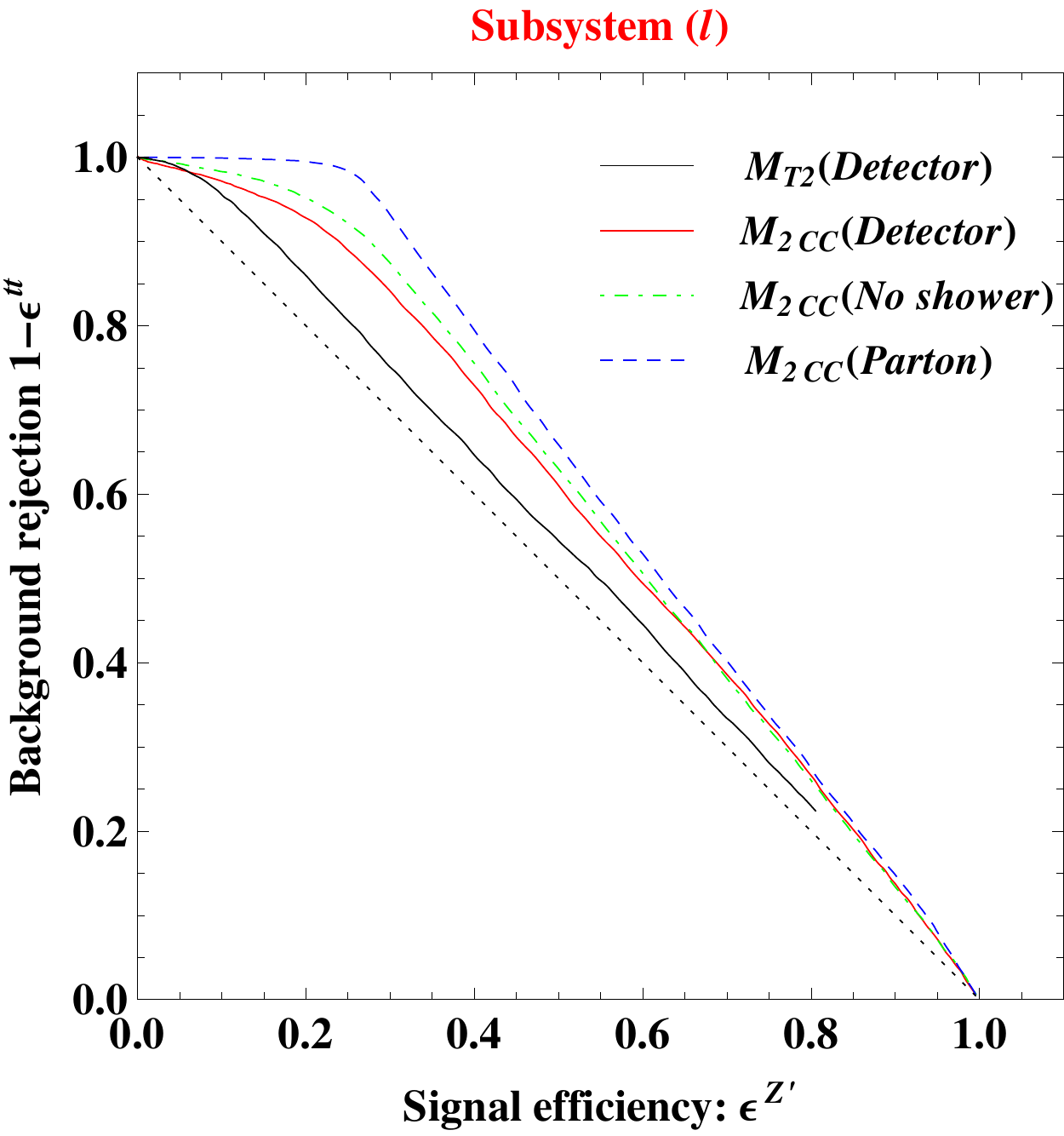}
\caption{\label{fig:CumRoc} Comparisons of signal and background efficiencies with $M_{2CC}$ and $M_{T2}$ variables for subsystems $(b\ell)$ (top panels), $(b)$ (middle panels), and $(\ell)$ (bottom panels). The chosen mass spectrum is that $m_{H^\pm} = 130$ GeV and $m_{Z'}=1$ GeV, and the test masses for the child particle are assumed to be $\tilde m = 0$ GeV for the $(b\ell)$ and $(\ell)$ subsystems and $\tilde m = 80$ GeV for the $(b)$ subsystem. The black dashed lines in the plots of the left panels denote the theoretical endpoints of $M_{T2}$ and $M_{2CC}$ for $t\bar{t}$. All curves are drawn with the detector-level events passing all selection cuts enumerated in (\ref{eq:cut1}) through (\ref{eq:cut4}). }
\end{figure}
%%%%%%%%%%%%%%%%%%%%%%

Our observations with the above standard variables strongly motivate alternative approaches to separate signal events from background ones. To begin with, the $M_{2CC}$ variable is contrasted with the standard $M_{T2}$ variable. When evaluating $M_{T2}$ and $M_{2CC}$, we make a use of the $M_2$ minimization code in Ref.~\cite{Cho:2015laa} and take the smaller one in the two possible combinations. We first remark that for the example spectrum under consideration, the $M_{T2}$ variable is not promising for the purpose of separating signal and background events. This is clearly shown in the Receiver Operating Characteristic (ROC) curves of the right panels in Figure~\ref{fig:CumRoc}, where the performance of $M_{T2}$ is described by the black curves. Since they all are below or closer to the diagonal line connecting $(1,0)$ and $(0,1)$ (black dotted lines in the right panels) than the associated curves for $M_{2CC}$, $M_{T2}$ is hardly beneficial or, at least, {\it not} the best option in selecting signal events against background ones. 

On the other hand, the red solid curves in Figure~\ref{fig:CumRoc} contrast the behavior of signal and $t\bar{t}$ events in $M_{2CC}$. We clearly observe that a larger fraction of signal events migrate to the regime beyond the expected kinematic endpoints of the $t\bar{t}$ system (vertical dashed lines) for the ($b\ell$) and ($\ell$) subsystems. Therefore, significant enhancement in the signal sensitivity is anticipated. In other words, setting $M_{2CC}$ cuts closer to (or even above) the expected kinematic endpoints for $t\bar{t}$ enables us to substantially suppress the background events while keeping a sizable number of signal events. In the later analysis, we do not use the $M_{2CC}$ for the $(b)$ subsystem because of its relatively poor performance. Thus we shall provide more detailed strategy combining the $M_{2CC}$ behaviors in the ($b\ell$) and ($\ell$) subsystems.

%%%%%%%%% FIGURE %%%%%%%%
\begin{figure}[t]
\centering
\includegraphics[width=7cm]{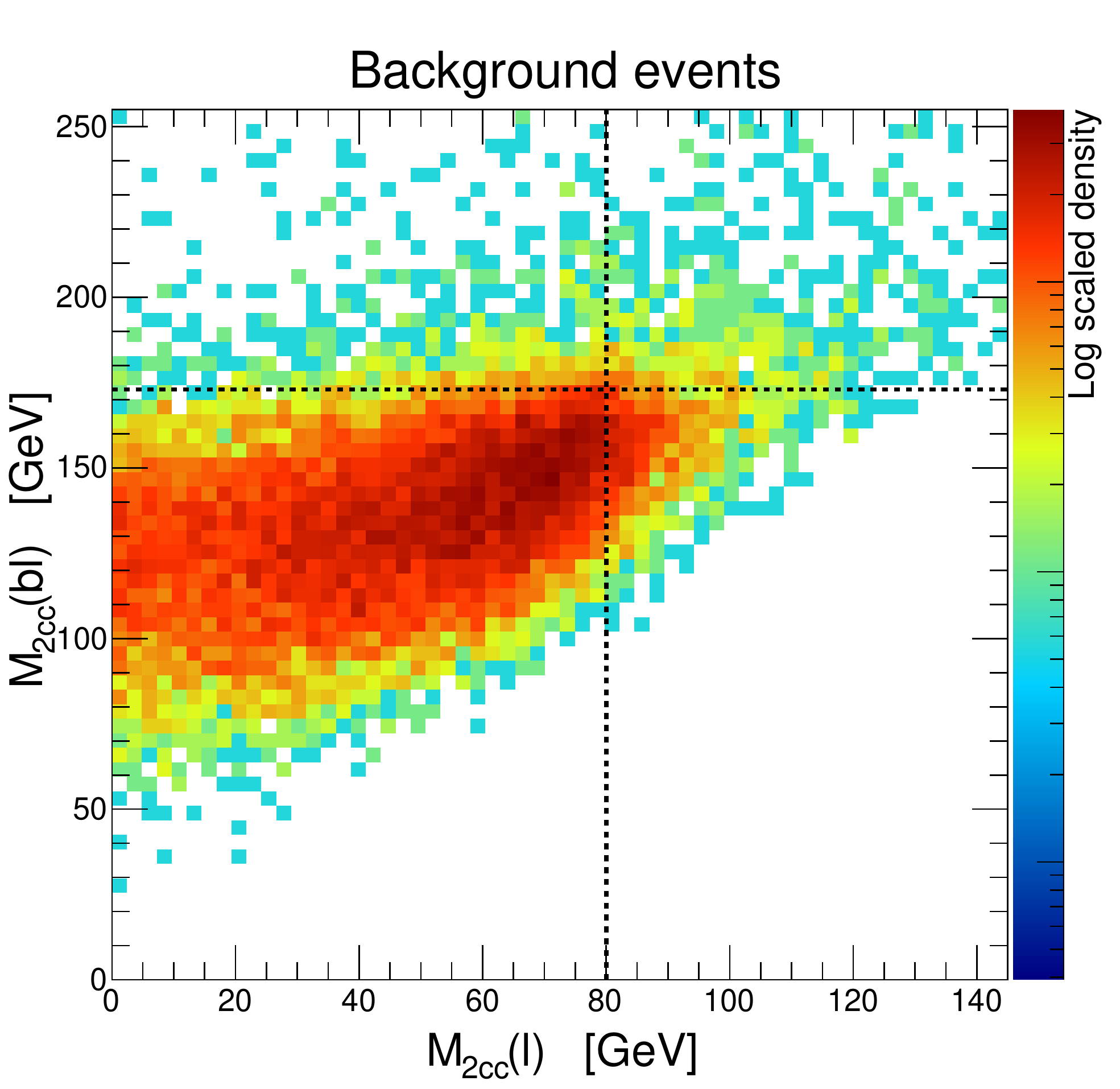} \qquad
\includegraphics[width=7cm]{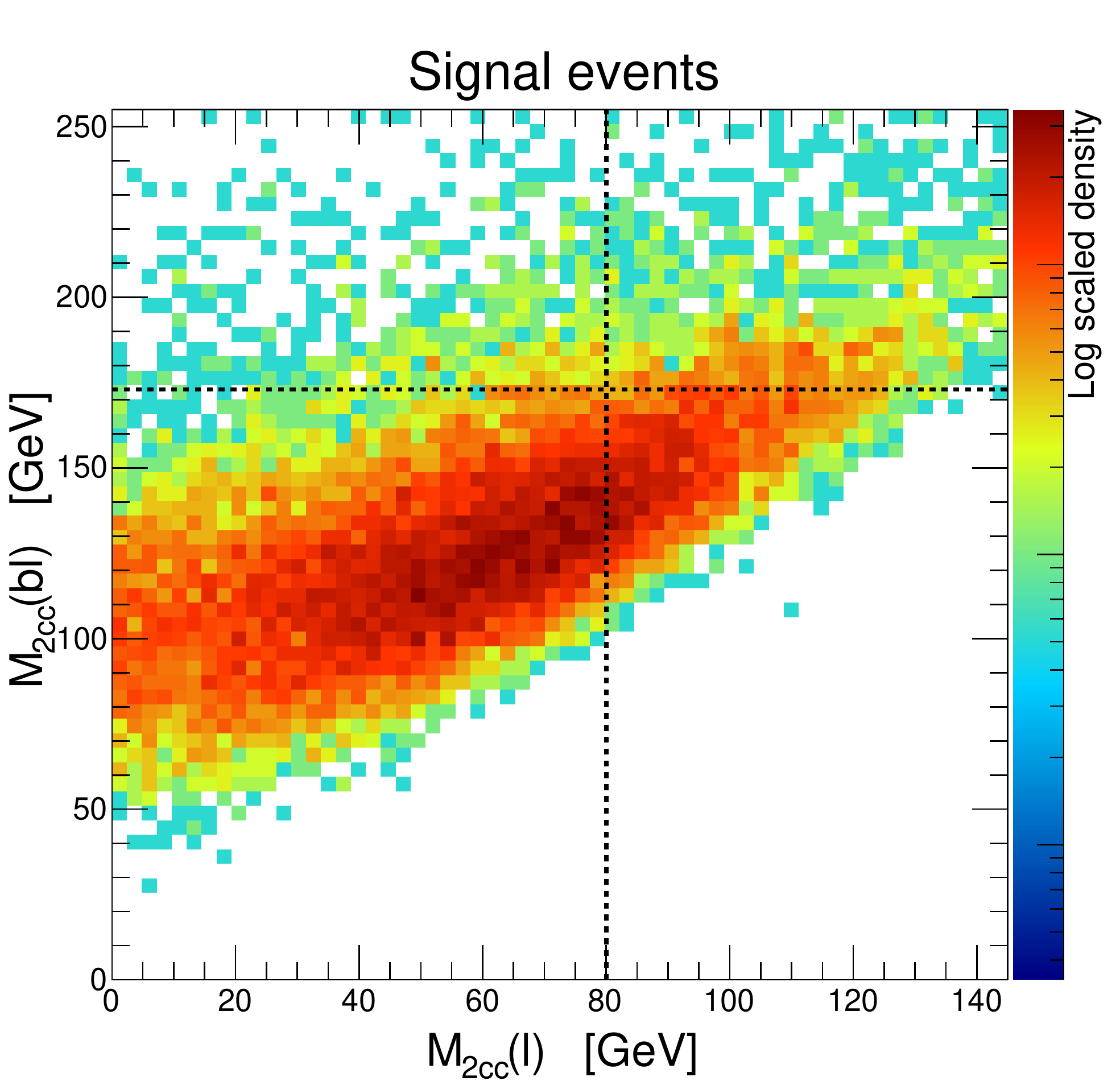}
\caption{\label{fig:2Dplot} Correlation plots of $M_{2CC}(b\ell)$ vs. $M_{2CC}(\ell)$ for background (left panel) and signal (right panel) events. For both subsystems, $\tilde m = 0$ GeV is imposed as the test mass, and the same study point as in Figure~\ref{fig:CumRoc} is chosen. Both temperature plots are produced with the detector-level events passing all selection cuts in (\ref{eq:cut1}) through (\ref{eq:cut4}). The dashed vertical and horizontal lines denote the theoretical endpoints of $M_{2CC}$ for $t\bar{t}$ in the respective subsystem. }
\end{figure}
%%%%%%%%%%%%%%%%%%%%%%

It actually deserves to check the correlation between the two $M_{2CC}$'s for the $(b\ell)$ and $(\ell)$ subsystems. Figure~\ref{fig:2Dplot} demonstrates the two-dimensional temperature plots of $M_{2CC}(b\ell)$ vs. $M_{2CC}(\ell)$. We see that the background events are inclined to populate towards the bottom-left corner (i.e., the third quadrant in the plane divided by the dashed vertical and horizontal lines). Based upon this observation, we separate the signal events from the $t\bar{t}$ events by the following posterior selection procedure:
\begin{itemize}
\item Given the two values $c_1$ and $c_2$, if $M_{2CC}$ values for any event satisfy either $M_{2CC}(b\ell)>c_1$ or $M_{2CC}(\ell)>c_2$, the event passes the test and is kept, and otherwise, it is rejected.
\end{itemize}
%
%%%%%%%%% TABLE %%%%%%%%%
\begin{table}[t]
\centering
\begin{tabular}{c|c c||c|c c| c c|| c}
  & $m_{H^{\pm}}$ & $m_{Z'}$ & $\mathcal{L}$ (fb$^{-1}$)  & $c_1$ & $c_2$ & $S\,(\times 10^3)$ & $B\,(\times 10^3)$ & $X\,(\times 10^{-3})$ \\
 \hline \hline 
 \multirow{2}{*}{BP1} & \multirow{2}{*}{130} & \multirow{2}{*}{1} & 300 & 360 & 90 & 1.10(0.44) & 48.1 & 11.0(4.4) \\ \cline{4-9}
 & & & 3000 & 305 & 90 & 3.49(1.40) & 485 & 3.5(1.4) \\
 \hline
  \multirow{2}{*}{BP2} & \multirow{2}{*}{130} & \multirow{2}{*}{20} & 300 & 330 & 92 & 1.04(0.42) & 43.3 & 12.5(5.0) \\ \cline{4-9}
 & & & 3000 & 285 & 92 & 3.32(1.33) & 440 & 4.0(1.6) \\
 \hline
  \multirow{2}{*}{BP3} & \multirow{2}{*}{130} & \multirow{2}{*}{5} & 300 & 330 & 90 & 1.10(0.44) & 48.3 &11.0(4.4) \\ \cline{4-9}
 & & & 3000 & 305 & 90 & 3.49(1.40) & 485 & 3.5(1.4) \\
 \hline
  \multirow{2}{*}{BP4} & \multirow{2}{*}{120} & \multirow{2}{*}{5} & 300 & 305 & 87 & 1.21(0.48) & 58.3 &17.3(6.9) \\ \cline{4-9}
 & & & 3000 & 285 & 87 & 3.83(1.53) & 587 & 5.5(2.2) \\
 \hline
  \multirow{2}{*}{BP5} & \multirow{2}{*}{110} & \multirow{2}{*}{5} & 300 & 360 & 87 & 1.21(0.48) & 57.9 &34.3(13.7) \\ \cline{4-9}
 & & & 3000 & 305 & 87 & 3.82(1.53) & 583 & 10.9(4.4)
\end{tabular}
\caption{\label{tab:discovery} $5\sigma$ discovery reach and $2\sigma$ exclusion limit (numbers in the parentheses) in $X = \br (t \to b W + Z')$ for several benchmark points with integrated luminosities of 300 fb$^{-1}$ and 3000 fb$^{-1}$ at $\sqrt{s}=14$ TeV LHC. $c_1$, $c_2$, and the masses of $H^\pm$ and $Z'$ are given in GeV. }
\end{table}
%%%%%%%%%%%%%%%%%%%%%%%%%
%
%%%%%%%%% TABLE %%%%%%%%%
%\begin{table}[t]
%\centering
%\begin{tabular}{c|c c c c c}
% & BP1 & BP2 & BP3 &BP4 &BP5\\
% \hline
%$m_{H^\pm}$ & 130 & 130 & 130 & 120 & 110 \\
%$m_{Z'}$        &1 & 20 & 5 & 5 & 5 \\
%\hline\hline
%$c_1$ (GeV) & 115\,(111) & 119\,(112) & 115\,(111) & 125\,(125) & 137\,(137) \\
%$c_2$ (GeV) & 79\,(79) & 83\,(77) & 88\,(80)  & 89\,(84) & 89\,(94)  \\
%\hline \hline
%$X\,(\times 10^{-3})$ & 6.7\,(2.2) & 6.5\,(2.1)  & 6.5\,(2.1) & 8.2\,(2.6) & 10.4\,(3.3) \\
%\end{tabular}
%\caption{\label{tab:discovery} $5\sigma$ C.L. discovery reach in $X = \br (t \to b W + Z')$ for several benchmark points with integrated luminosities of 300 fb$^{-1}$ (3000 fb$^{-1}$) at $\sqrt{s}=14$ TeV LHC. The $H^\pm$ and $Z'$ masses are given in GeV.}
%\end{table}
%%%%%%%%%%%%%%%%%%%%%%
%
For a given study point, we basically vary $c_1$ and $c_2$ to find the best combinations which can give rise to statistical significances $S/\sigma_B$ of $\sim5\sigma$ and $\sim 2\sigma$, where $S$ is the expected number of signal events while $\sigma_B$ is taken as the Gaussian approximate of Poisson statistical errors, i.e., $\sigma_B = \sqrt{B}$ with $B$ being the expected number of background events. Table~\ref{tab:discovery} lists five selective benchmark points that are investigated in our simulations and the expected reaches of the branching ratio $X = \br (t \to b W + Z'\text{s})$ for $5\sigma$ and $2\sigma$ (numbers in the parentheses) excesses are provided under integrated luminosities of 300 fb$^{-1}$ and 3000 fb$^{-1}$ with the total center of mass energy being 14 TeV, accompanying the optimized set of $c_1$ and $c_2$. We do not consider the mass of the charged Higgs heavier than 130 GeV for the following reason. The small mass gap between the top quark and the charged Higgs causes a soft emission of the bottom quark so that less signal events are likely to pass our selection criteria. This implies that a significant deficiency from the SM $t\bar{t}$ cross section would have been observed even before applying $M_{2CC}$ cuts.\footnote{In principle, it is possible to discover $Z'$ by observing a significant deficiency from the pure SM prediction even in conjunction with $M_{2CC}$ cuts. This approach is interesting {\it per se}, but we do simply pursue the conventional direction in this paper.}

From our simulation study with the above-given benchmark points, we make a couple of observations about the sensitivity of the proposed technique to $m_{H^{\pm}}$ and $m_{Z'}$. First, the closer the mass of the charged Higgs is to the mass of the $W$ gauge boson, the less effective the technique is (see BP3 through BP5). As an extreme case, if $m_{H^{\pm}}$ became nearly degenerate to $m_W$, then $Z'$ would become extremely soft so that the overall ensemble in the final state would get similar to the regular top decay. In other words, the topologically asymmetric nature of the signal process effectively disappears.  
Second, our technique is less sensitive to the mass of $Z'$ unlike the case of $m_{H^{\pm}}$ (see BP1 through BP3). Typically, the details of $b$ and $\ell$ are determined by the $t - H^{\pm}$ and $W -\nu$ mass gaps, correspondingly. Therefore, the details of the $Z'$ mass do not make any significant effect on the overall ensemble in the visible state. This observation is actually good for the signal channels having multiple $Z'$s via an (on-shell) non-SM, light scalar since, for example, a large parameter space with $\br(h \rightarrow Z'Z') \sim 1$ \cite{Lee:2013fda} can be accommodated in replacing $Z'$ with $h$. Regarding the potential spin sensitivity of our technique, we also perform similar exercises with parton-level event samples involving $Z'$ or $h$, and find that the distributions in $M_2$ variables for $h$ are almost the same as those for $Z'$, i.e., negligibly sensitive to the detailed spin assignment in the associated decay sequence (see also Ref.~\cite{Edelhauser:2012xb} for the $M_{T2}$ variable). Therefore, we expect effectively the same signal efficiency (or background veto) in the signal process having $h$. 

We point out that unlike the theory prediction, which is also supported by the parton level simulation in Figure~\ref{fig:partonlevel}, even some fraction of $t\bar{t}$ events show an endpoint violation in the detector level simulation. Two possible sources can be taken into account. First, the initial and final state radiations (ISR/FSR) can cause such a phenomenon. For example, if one of the two $b$ quarks is too soft to form a jet while an ISR/FSR jet is mis-tagged, then the resultant decay topology is ill-defined, i.e., contradictory to the model assumptions in $M_{2CC}$, so that the corresponding $M_{2CC}$ value could end up with being beyond the expected kinematic endpoint. To verify this argument, we perform the simulation with the ISR and FSR turned off (green curves in Figure~\ref{fig:CumRoc}). Not surprisingly, the signal efficiency does not get noticeably altered because the endpoint violation comes dominantly from the asymmetric nature of the signal process. On the contrary, it is shown that the $t\bar{t}$ events beyond the expected kinematic endpoint are pushed to the left, and as a consequence, the background rejection above the kinematic endpoint becomes better while the signal efficiency is retained. We therefore expect that proper handling of the ISR/FSR jets will improve the performance of our search strategy. The other cause can be the mis-measurement of jets and missing transverse momentum.\footnote{Here we assume that the experimental quantities for the leptons are much better-measured than the jets.} To see such effect, we compare the parton-level results\footnote{Only the non-zero decay width is in effect. 
} described by the blue curves in Figure~\ref{fig:CumRoc} with the green curves (i.e., no ISR/FSR). Depending on the subsystems, the improvement is also recognizable or even better than the corresponding improvement in the case without ISR/FSR. Consequently, it is expected that a better understanding in the jet energy/missing energy resolution will sharpen the topological difference between the signal and background events.   

Finally, we briefly discuss the impact of the inclusion of systematics on the discovery potential of $Z'$. Due to large statistics in the search channel of interest, the final uncertainty tends to be dominated by the systematic uncertainty (denoted by $\sigma_B^{\textnormal{sys}}$), and thus given a moderate $\sigma_B^{\textnormal{sys}}\sim \mathcal{O}(20\%)$, the signal will be easily overwhelmed by the resulting uncertainty. It is therefore important to have the relevant systematics well under control in order to make our search strategy feasible. In general, identifying all sources and estimating the associated errors are highly non-trivial. Instead of pursuing such a direction, we rather take the corresponding number from the experiments looking at the same channel with comparable statistics as a reference. For example, Ref.~\cite{CMS} has reported an uncertainty of $\mathcal{O}(3\%)$ in the expected number of dileptonic $t\bar{t}$ events. We perform a similar analysis as in Table~\ref{tab:discovery} with the relevant significance $(\sigma)$ modified as
\bea
\sigma = \frac{S}{\sqrt{B+\left(\frac{\sigma_B^{\textnormal{sys}}}{100\%} \right)^2B^2}}, \label{eq:modsig}
\eea
and show, in Table~\ref{tab:discWsys}, $5\sigma$ and $2\sigma$ (numbers in the parentheses) reaches of the branching fraction $X$ for BP1 under an integrated luminosity of 300 fb$^{-1}$ with $\sqrt{s}=14$ TeV according to a few values of $\sigma_B^{\textnormal{sys}}$. We do not report the numbers for the case where the branching fraction $X$ is too large, say $10\%$. Clearly, this analysis suggests that there be still discovery opportunity of $Z'$ with a decent $\sigma_B^{\textnormal{sys}}$, and even with severe systematics we might have a mild excess if the relevant branching fraction is sizable enough. 
\begin{table}[t]
\centering
\begin{tabular}{c|c c c c}
$\sigma_B^{\textnormal{sys}}$ (\%) & 1 & 3 & 5 & 10 \\
\hline \hline
%$c_1$ & 363 (361) & 363 (361) & 363 (363) & --\;  (363) \\
%$c_2$ & 108 (108) & 108 (108) & 108 (108) & --\;  (108)\\
%\hline
$S(\times 10^3)$ & 0.95 (0.38) & 2.85 (1.14) & 4.74 (1.90) & --\;\;  (3.80)\\
$B(\times 10^3)$ & 19.0 & 19.0 & 19.0 & 19.0 \\
\hline 
$X(\times 10^{-3})$ & 18.1 (7.3) & 54.2 (21.7) & 90.3 (36.2) & --\;\;  (72.3)
\end{tabular}
\caption{\label{tab:discWsys} $5\sigma$ discovery reach and $2\sigma$ exclusion limit (numbers in the parentheses) in $X = \br (t \to b W + Z')$ for BP1 with an integrated luminosity of 300 fb$^{-1}$ at $\sqrt{s}=14$ TeV LHC according to a few different values of $\sigma_B^{\textnormal{sys}}$. The best combination of $c_1$ and $c_2$ is 360 GeV and 108 GeV, respectively. The associated numbers for the case with $X>0.1$ are not reported.  }
\end{table}

One potential confusion arising in comparing Tables~\ref{tab:discovery} and~\ref{tab:discWsys} is that they have different signal sensitivities, stemming from the introduction of a different set of cuts; the analysis with systematics applies more severe cuts. For the systematics-dominated data sample containing a sizable number of background events, the significance in Eq.~(\ref{eq:modsig}) can be approximated to $\sigma \approx \frac{S}{(\sigma_B^{\textnormal{sys}}/100\%)B}$ so that the application of the cuts in Table~\ref{tab:discovery} would have resulted in poor significance. To resolve this issue, one could either increase the number of signal events by increasing the signal branching fraction $X$ or suppressing more background events using harder cuts. We have tried both approaches to get the best reach, i.e., the smallest value of $X$, and found that a suitable combination between them enables us to have the best probe into the branching ratio $X$. 

%%%%%%%%%%%%%%%%%%%%%%%%%%%%%%%%%%%%%%%%%
%%%%%%%%%%%%%%%%%%%%%%%%%%%%%%%%%%%%%%%%%
\section{\label{sec:conclusion} Conclusions} 
%%%%%%%%%%%%%%%%%%%%%%%%%%%%%%%%%%%%%%%%%
%%%%%%%%%%%%%%%%%%%%%%%%%%%%%%%%%%%%%%%%%
In this work, we studied the discovery opportunity of a dark force mediator $Z'$ at the 14 TeV LHC. $Z'$ here is assumed invisible, and thus our study can be taken as complementary to Ref.~\cite{Kong:2014jwa} where $Z'$ is assumed to decay visibly. The chosen signal process is defined as a rare decay of top quark. Considering the facts that the production cross section for top quark pairs is huge at the LHC and the decay width of top quark is less precisely measured than that of other SM particles, it is clear that the signal process at hand can serve as a great discovery channel. 

On the other hand, by construction, $t\bar{t}$ itself plays a role of the dominant irreducible background to the $Z'$ signal at the same time so that the relevant searches are typically very challenging again due to its vast production cross section. To get over this difficulty and increase the associated signal sensitivity, we employed the recently proposed on-shell constrained $M_2$ variables as a tool. Among those kinematic variables, we chose $M_{2CC}$ that is constructed under the assumption of dileptonic $t\bar{t}$-like event topology. Since the signal process comes with a different decay topology from $t\bar{t}$, it is expected that applying $M_{2CC}$ to the signal events tends to give rise to some contradictory result. One possible observable visualizing such a contradiction is the substantial departure from the kinematic endpoint predicted in the context of the dileptonic $t\bar{t}$ system. 

To see the viability of our technique, we performed a Monte Carlo study including realistic effects such as cuts and detector resolutions. Depending on the masses of $H^{\pm}$ and $Z'$, the reach for the effective branching fraction $X$ defined in Eq.~(\ref{eq:Xdef}) can be $\sim1.1$\% ($\sim0.35$\%) under an integrated luminosity of 300 fb$^{-1}$ (3000 fb$^{-1}$) together with a suitable set of $M_{2CC}$ cuts.  We also pointed out that the proposed method can be generic enough to apply the main idea for the models where $Z'$ is replaced by another new particle with a different spin, e.g., a non-SM scalar ($h$).  Finally, since the signal process of interest involves $b$-tagged jets, the potential of improving our technique was discussed in the aspect of handling the ISR/FSR and the jet energy resolution.

\begin{acknowledgments}
We thank K.~T.~Matchev for useful comments and discussions. HL thanks H.~Davoudiasl and W.~Marciano for long-term discussions on the physics of dark gauge boson. DK is supported by the LHC Theory Initiative postdoctoral fellowship (NSF Grant No. PHY-0969510), and HL is supported by the CERN-Korea fellowship. MP was supported by World Premier International Research Center Initiative (WPI), MEXT, Japan. MP also acknowledges the Max-Planck-Gesellschaft, the Korea Ministry of Education, Science and Technology, Gyeongsangbuk-Do and Pohang City for the support of the Independent Junior Research Group at the APCTP.
\end{acknowledgments}

%%%%%%%%%%%%%%%%%%%%%%%%%%%%%%%%%%%%%%%%%
%%%%%%%%%%%%%%%%%%%%%%%%%%%%%%%%%%%%%%%%%

\end{document}